\documentclass[suppldata]{interact}

\usepackage{subfigure}
\usepackage{comment}
\usepackage{longtable}
\usepackage{afterpage} 
\usepackage{pdflscape}
\usepackage[colorinlistoftodos,prependcaption,textsize=tiny]{todonotes}
\usepackage{natbib}
\bibpunct[, ]{(}{)}{;}{a}{}{,}
\renewcommand\bibfont{\fontsize{10}{12}\selectfont}

\usepackage[font=normalsize]{caption} 
\theoremstyle{plain}

\theoremstyle{definition}

\theoremstyle{remark}

\usepackage{multirow}

\usepackage{array}
\usepackage{svg}
\svgpath{{images_pdf/}} 
\makeatletter
\newcommand{\thickhline}{%
	\noalign {\ifnum 0=`}\fi \hrule height 1.5pt
	\futurelet \reserved@a \@xhline
}
\newcolumntype{"}{@{\hskip\tabcolsep\vrule width 1pt\hskip\tabcolsep}}
\makeatother
\usepackage{soul}

\usepackage{threeparttable}
\usepackage{graphicx}
\usepackage{color,soul}
\usepackage{indentfirst}

\usepackage{hyphenat}
\begin{document}

	\title{DeepVol: Volatility Forecasting from High-Frequency Data with Dilated Causal Convolutions}
	
	\author{
		\name{Fernando~Moreno-Pino\textsuperscript{a,b}\thanks{Corresponding author: fernando.moreno-pino@eng.ox.ac.uk} and Stefan~Zohren\textsuperscript{a,c}}
		\centering
		\affil{\textsuperscript{a}Oxford-Man Institute of Quantitative Finance, University of Oxford\\
			\textsuperscript{b}Signal Processing and Learning Group, Universidad Carlos III de Madrid\\
			\textsuperscript{c}Machine Learning Research Group, University of Oxford}}
	
	\maketitle

	\begin{abstract}
		Volatility forecasts play a central role among equity risk measures. Besides traditional statistical models, modern forecasting techniques based on machine learning can be employed when treating volatility as a univariate, daily time-series. \textcolor{black}{Moreover}, econometric studies have shown that increasing the number of daily observations with high-frequency intraday data helps to improve volatility predictions. In this work, we propose DeepVol, a model based on Dilated Causal Convolutions that uses high-frequency data to forecast day-ahead volatility. 
		\textcolor{black}{Our empirical findings demonstrate} that dilated convolutional filters are \textcolor{black}{highly effective at extracting} relevant information from intraday financial \textcolor{black}{time-series}, \textcolor{black}{proving that this architecture can effectively leverage predictive information present in high-frequency data that would otherwise be lost if realised measures were precomputed.}
		Simultaneously, dilated convolutional filters trained with intraday high-frequency data help us \textcolor{black}{avoid} the limitations of models that use daily data, such as model misspecification or manually designed handcrafted features, whose devise involves optimising the trade-off between accuracy and computational efficiency and makes models prone to lack of adaptation into changing circumstances.
		In our analysis, we use two years of intraday data from NASDAQ-100 to evaluate the performance of DeepVol.
		Our empirical results suggest that the proposed deep learning-based approach \textcolor{black}{effectively} learns global features from high-frequency data, \textcolor{black}{resulting in} more accurate predictions \textcolor{black}{compared to} traditional methodologies and producing more \textcolor{black}{accurate} risk measures.
	\end{abstract}


	\section{Introduction } 
	\label{sec:introduction}

	\begin{figure}[]
		\centering
		\includegraphics[width=12cm,height=10cm,keepaspectratio]{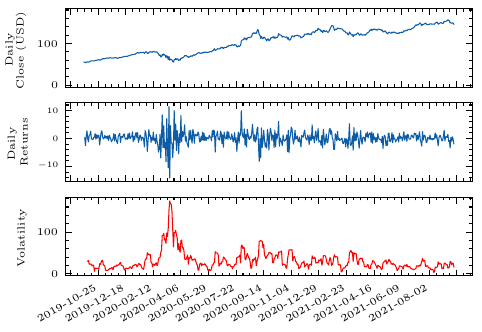}
		\caption{Apple's daily data. \textcolor{black}{ The top row shows the price trend, while the second row depicts the associated daily returns. In the bottom row, an estimation of the unobserved latent volatility process is calculated using a 5-day moving window over the daily returns. Notice that this method of estimating the volatility serves just as an approximation to the underlying latent volatility process, providing insights into the dynamic nature of market volatility.}}
		\label{fig:data_example}
	\end{figure}

	In recent years, \textcolor{black}{the assessment of portfolios' risk through volatility measures has garnered significant attention \citep{brownlees2010comparison},  leading to the growing adoption} of volatility conditional portfolios \citep{harvey2018impact}\textcolor{black}{. Various studies have reported  overall gains in their Sharpe ratios}  \citep{moreira2017volatility} \textcolor{black}{ and reductions in the likelihood of observing extreme heavy-tailed returns} when using them \citep{harvey2018impact}. \textcolor{black}{Consequently, there has been widespread interest in the development of volatility forecasting models.}\\
	
	\textcolor{black}{In volatility forecasting, as in many other forecasting problems in economics and finance, the variable of interest is latent. This is exemplified in Figure \ref{fig:data_example}, where we employ a rolling window to estimate the unobserved latent volatility process from the observed returns. To address this complexity, an unbiased estimator of this latent variable must be chosen as a proxy measure. Within the scope of volatility forecasting, the squared returns of an asset over a specific period of time is one of the most obvious realised volatility measures that can be interpreted as a conditionally unbiased estimator of the true unobserved conditional variance of the asset over that same time span \citep{patton2011volatility}.}\\

	\textcolor{black}{In a context where} volatility forecasting methods \textcolor{black}{offer considerable advantages, 
		with an extensive literature also pointing out that intraday volatility forecasts are important for improving the understanding of the risk involved in trading strategies \citep{bates2019crashes}, pricing derivatives and the development of quantitative strategies \citep{engle2012forecasting}, as well as for risk management and trading applications \citep{stroud2014bayesian}, among others,
		numerous studies have explored the challenges of intraday and daily volatility prediction.  Delving into the latter, practitioners commonly rely on classic methodologies, with a heavy reliance on models that provide forecasts of daily volatility from daily returns.  The Generalized Autoregressive Conditional Heteroskedasticity (GARCH) model \citep{bollerslev1986generalized} and traditional stochastic volatility models \citep{andersen2006volatility, calvet2006volatility} are predominant in this regard. These models leverage volatility spillover effects and past volatility along with daily squared returns as the driving variables for predicting day-ahead volatility. 
		However, their effectiveness is constrained by their inability to effectively leverage high-frequency data \citep{andersen2003modeling, andersen2006volatility, engle2007good} and their inefficacy  to  analysing simultaneously multiple assets in high-dimensionality.}\\

	\textcolor{black}{Recent advancements have addressed some of these challenges by incorporating} realised measures as predictors for realised volatility\textcolor{black}{, thereby enhancing the prediction accuracy of classic models \citep{hansen2012realised, patton2022bespoke}. Realised measures}, non-parametric estimators of an asset price variation \textcolor{black}{over a time interval,} extract and summarize information \textcolor{black}{embedded} in high-frequency data \citep{andersen2010parametric}\textcolor{black}{. Moreover, ongoing research continues to introduce novel methodologies for computing realised measures. E.g., \citet{pascalau2023increasing} proposes a new approach to estimate and forecast realised volatility that contrast to the usual calendar approach.
	}\\
	
	However, methodologies that take advantage of realised measures require pre-processing steps to use them, as they cannot directly model the complex relations exhibited by intraday financial data. One concrete example of this pre-processing is the procedure followed to obtain the realised measures themselves, which summarises vectors of daily intraday high-frequency data into single scalars \textcolor{black}{to avoid managing} the microstructure noise associated with the former.
	In contrast, \textcolor{black}{our approach} uses raw high-frequency data as input to the model, \textcolor{black}{eliminating the need for data preprocessing and mitigating associated drawbacks}, such as \textcolor{black}{information loss during the aggregation of intraday data into daily realised measures}. Our proposed model is \textcolor{black}{capable} of directly handling the microstructure noise linked to higher intraday data's sampling frequencies.\\

	Among the methodologies employing realised measures, the HEAVY model \citep{shephard2010realising} is of special appeal among industry practitioners \citep{karanasos2022emerging, papantonis2022improving, yuan2022high}. HEAVY is based on insights from the ARCH architecture, with superior performance over other classical benchmarks, as shown in Section \ref{sec:experiments}.
	Nevertheless, the inability of realised measures-based models, such as HEAVY\textcolor{black}{, Realised GARCH \citep{hansen2012realised}, or HAR \citep{corsi2009simple}} to use unprocessed raw high-frequency data \textcolor{black}{directly} as input, exposes them to several disadvantages. Firstly, the dependence on the realised measures for day-ahead volatility forecasting artificially limits the amount of information these architectures use, which is not the case when using raw intraday data\textcolor{black}{, as proven in Section \ref{sec:experiments}}. Furthermore, some of the most used realised measures of volatility lack robustness to microstructure noise \citep{baars2014heavy}, implying that the trained models may be based on biased data. Finally, methodologies based on realised measures often rely on manually designed handcrafted features, as the realised measures design \textcolor{black}{themselves}, formulated to optimise the trade-off between accuracy and increasing computational costs. \textcolor{black}{These issues, coupled} with common model misspecification \textcolor{black}{in} classical model-based approaches, undermines their reported performances.\\

	Here, we use Deep Neural Networks (DNN) \textcolor{black}{\citep{lecun2015deep}} to take advantage of the abundance of high-frequency data without prejudice, preventing the constraints of models based on realised measures in the context of day-ahead volatility forecasting.
	Despite the success of these DNN architectures in different areas, such as healthcare \textcolor{black}{\citep{shamshirband2021review}}, image recognition \textcolor{black}{\citep{kaur2020deep, chen2021review}}, \textcolor{black}{ time-series modelling and forecasting \citep{li2019enhancing, jimenez2023interpretable, moreno2024rough}, }and text analytics \textcolor{black}{\citep{conneau2016very}}, they have not been widely adopted for the problem of volatility forecasting, leading to a large gap between modern machine learning models and those applied in the volatility framework. Among DNN-based models, Recurrent Neural Networks (RNN)  \citep {rumelhart1985learning} and Long Short-Term Memory (LSTM) \citep{hochreiter1997long} are the most popular approaches with regard to time-series forecasting \citep{lim2021time}. Furthermore, the addition of the attention mechanism \citep{bahdanau2014neural} into these base architectures allowed them to focus on the most relevant input data while producing predictions, making them especially prominent in fields such as Natural Language Processing (NLP). These advances also lead to the appearance of  Transformer models \citep{vaswani2017attention}, which were initially introduced for NLP, and later used for the problem of time-series forecasting \citep{li2019enhancing, moreno2021deep}.
	These models 
	are \textcolor{black}{also} applied in the context of financial time-series through diverse variations \textcolor{black}{\citep{su2021promote, scalzo2021nonstationary, lin2022forecasting, arroyo2022dynamic, christensen2023machine, zhu2023forecasting, song2023volatility, arroyo2024deep, souto2024introducing, ghosh2024deep}}. More specifically, regarding volatility forecasting, \textcolor{black}{various studies have delved into the application} of deep-learning architectures,
	such as LSTM \citep{yu2018forecasting}, Convolutional Neural Networks (CNN) \citep{vidal2020gold},  Graph Neural Networks (GNN) \textcolor{black}{\citep{chen2021multivariate, zhang2022volatility}}, Transformer models \textcolor{black}{\citep{ramos2021multi}}, and NLP-based word embedding techniques \citep{rahimikia2020big, rahimikia2021realised}. Furthermore, models combining traditional volatility forecasting methods with deep-learning techniques can be found in the literature \citep{kim2018forecasting, mademlis2021volatility}, as well as other approaches using DNN as calibration methods for implying volatility surfaces \citep{horvath2019deep}, proving how neural network-based approaches work as complex pricing function approximators.\\

	We capitalise on the increased availability of high-frequency data. In this work\textcolor{black}{, }we employ a Dilated Causal Convolutions (DCC)-based model. This architecture, initially proposed as a fully probabilistic model for audio generation \citep{oord2016wavenet}, with equivalents for image-related problems \citep{van2016pixel}, \textcolor{black}{can handle lengthy sequences of data effectively} \textcolor{black}{without a significant increase in the number of parameters} \textcolor{black}{thanks to the use of dilated connections.}
	In the literature, other works use DCC in the context of realised volatility forecasting. 
	More specifically, \citet{reisenhofer2022harnet} propose a model based on dilated convolutions, strongly inspired by the well-known Heterogeneous Autoregressive (HAR) model \citep{corsi2009simple}.
	However, their approach does not use unprocessed raw intraday high-frequency data as input. Conversely, it still bases its predictions on the pre-computed daily realised variance, therefore requiring pre-processing steps to obtain the indispensable realised measures for forecasting the one-step-ahead volatility. This, in our judgment, does not fully explore the capabilities of DCC-based methodologies of exploiting a more dynamic representation of the intraday data. 
	Hence, models adopting DCC-based approaches that operate from daily data still succumb to the limitations enumerated previously.\\

	Motivated  by the improved performance of classical methods that employ realised measures \citep{hansen2012realised, shephard2010realising}, we propose using DCCs to bypass the estimation of these non-parametric estimators of assets' variance, aiming to tackle the volatility forecasting problem from a data-driven perspective. The proposed model, DeepVol, \textcolor{black}{offers} several advantages \textcolor{black}{for volatility forecasting}. Primarily, it does not require any pre-processing steps, as the model directly uses raw high-frequency data as input. Furthermore, DeepVol is not bounded to static realised measures whose use may be counter-productive, i.e. the optimal realised measure to use may vary depending on the traded assets' liquidity. 
	Instead, through the attention mechanism and internal non-linearities, DeepVol intelligently performs the required transformations over the input data to maximise the accuracy of the predictions, combining relevant intraday datapoints and merging them for each day's volatility forecast, dinamically adapting to different scenarios.
	Moreover, through the use of dilated convolutions, DeepVol's large receptive field easily processes long sequences of high-frequency data, enabling the model to exponentially increase its input window \textcolor{black}{without triggering an unrestrained increase in the model’s complexity} while performing the predictions. 
	\textcolor{black}{
		While DeepVol's operation does not directly produce any type of realised measure ---meaning the model does not explicitly construct an ex-post estimate of the returns variation--- its aggregation of high-frequency data resembles how realised measures condense intraday data into daily statistics.
		Unlike the computation of these realised measures, DeepVol undertakes this aggregation in a data-driven manner, diverging from traditional approaches used for computing realised measures of volatility. 
		DeepVol's ability to dynamically select and weight the most relevant time-steps from the conditioning range at each time instant stands out as its primary distinction from traditional methods computing realised measures, allowing DeepVol to hierarchically integrate the most relevant high-frequency data into the predictions.}
	We perform extensive experiments to show the effectiveness of the proposed architecture, which consistently outperforms the base models used by practitioners.\\

	This paper \textcolor{black}{presents} three main contributions. Firstly, \textcolor{black}{it} empirically \textcolor{black}{demonstrates} the advantages \textcolor{black}{of DCCs for forecasting realised volatility using high-frequency data}, providing a data-driven solution \textcolor{black}{that} consistently outperforms classical methodologies. \textcolor{black}{Further,} the proposed model \textcolor{black}{overcomes} the limitations of classical methods, such as model misspecification or their inability to directly use intraday data to perform the forecast.
	Secondly, we \textcolor{black}{offer} an analysis \textcolor{black}{showing how DeepVol} optimizes the balance between extracting signals from high-frequency data and minimizing the microstructure noise inherent \textcolor{black}{to} higher sampling frequencies. 
	The reported results are consistent with studies validating this trade-off for constructing realised measures.
	Thirdly, the proposed volatility forecasting model generates appropriate risk measures through its predictions in an out-of-sample forecasting task, both in low and high volatility regimes. Moreover, we evaluate the proposed model's generalisation capabilities on out-of-distribution stocks, demonstrating DeepVol's capabilities to transfer learning as it performs accurate predictions into data distributions not observed during the training phase.\\

	The structure of the paper is as follows. Section \ref{sec:data_inputs} details the dataset used, while Section \ref{sec:baselines_metrics} contains a brief overview on volatility forecasting, describing the baselines used for benchmarking purposes and the metrics that will be utilised for model comparison. Section \ref{sec:model} presents the proposed model, which is empirically evaluated in Section \ref{sec:experiments}. Finally, Section \ref{sec:conclusions} summarises the findings and \textcolor{black}{provides concluding remarks}.\\
	

	\section{Data and Model Inputs}
	\label{sec:data_inputs}
	
	\subsection{Data}
	\label{subsec:data}
	
	\begin{figure}[t!]
		\includegraphics[width=0.5\columnwidth]{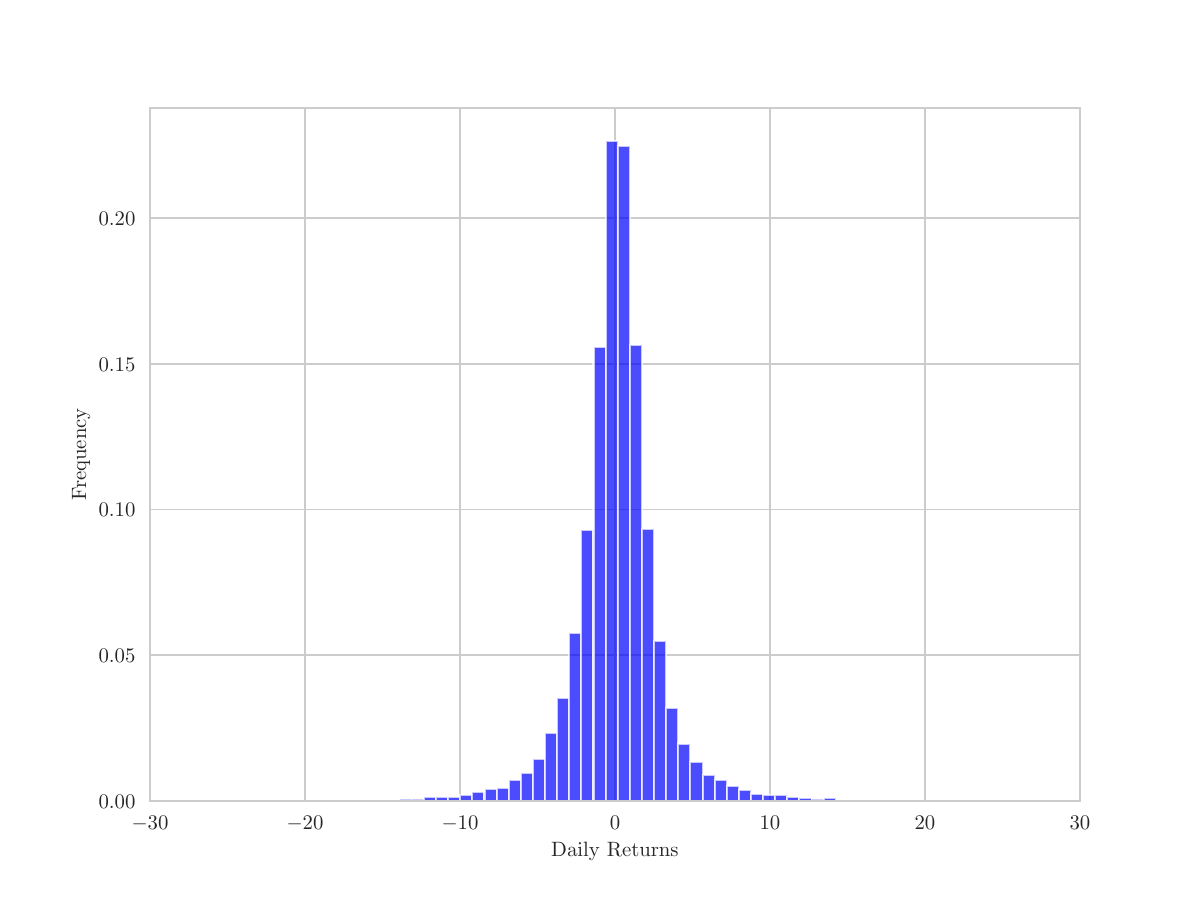}
		\hspace*{\fill}
		\includegraphics[width=0.5\columnwidth]{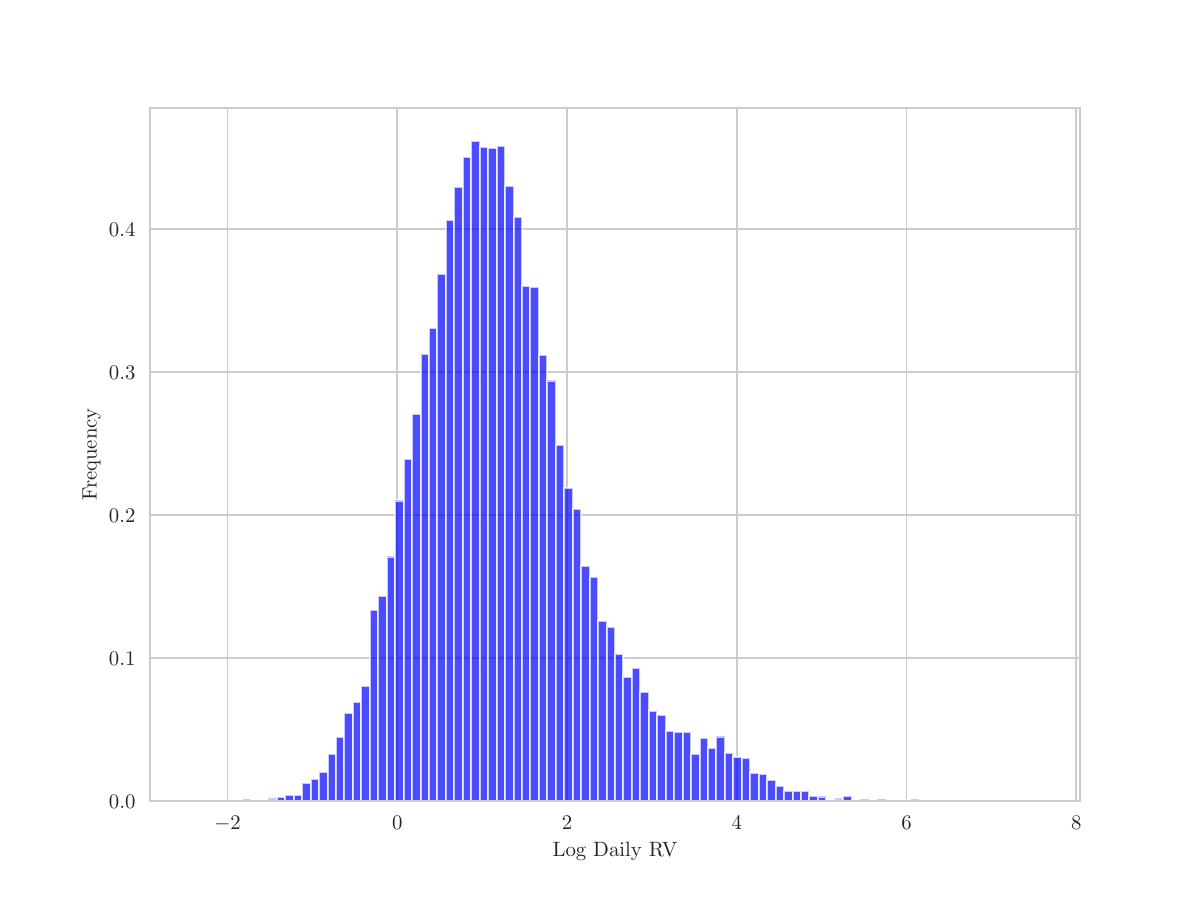}
		\caption{\textcolor{black}{\textbf{Left: } Distribution of daily returns. \textbf{ Right: } Distribution of realised volatility (in logs).}}
		\label{fig:hists}
	\end{figure}

\begin{figure}[t!]
	\begin{center}
		\centerline{\includegraphics[width=1.1\columnwidth]{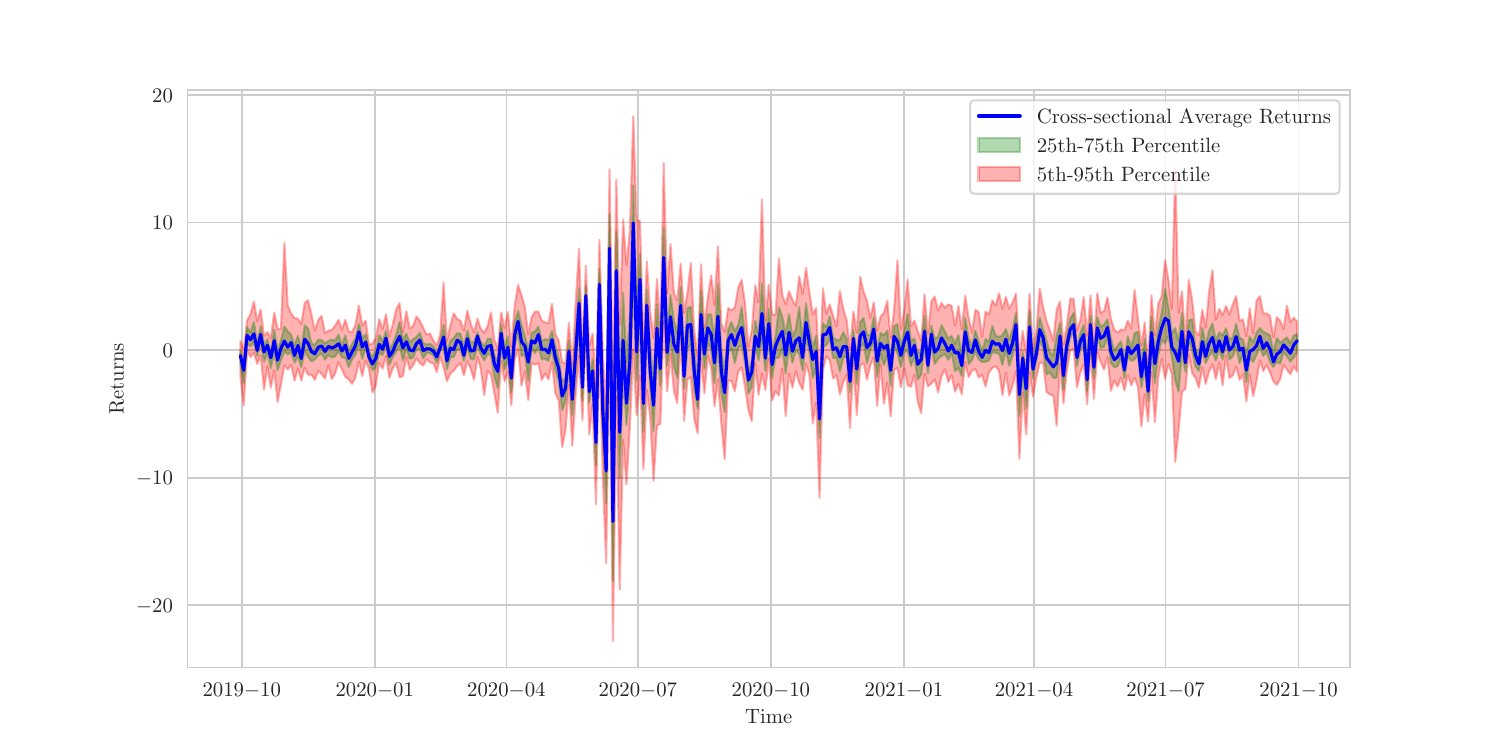}}
		\caption{\textcolor{black}{The blue curve represents \textcolor{black}{the} cross-sectional average of daily returns across the analysed stocks, with the green area covering the 25-th percentile to the 75-th percentile, and the red area covering the 5-th percentile to the 95-th percentile.}} 
		\label{fig:percentiles_rt}
	\end{center}
\end{figure}

\begin{figure}[t!]
	\begin{center}
		\centerline{\includegraphics[width=1.1\columnwidth]{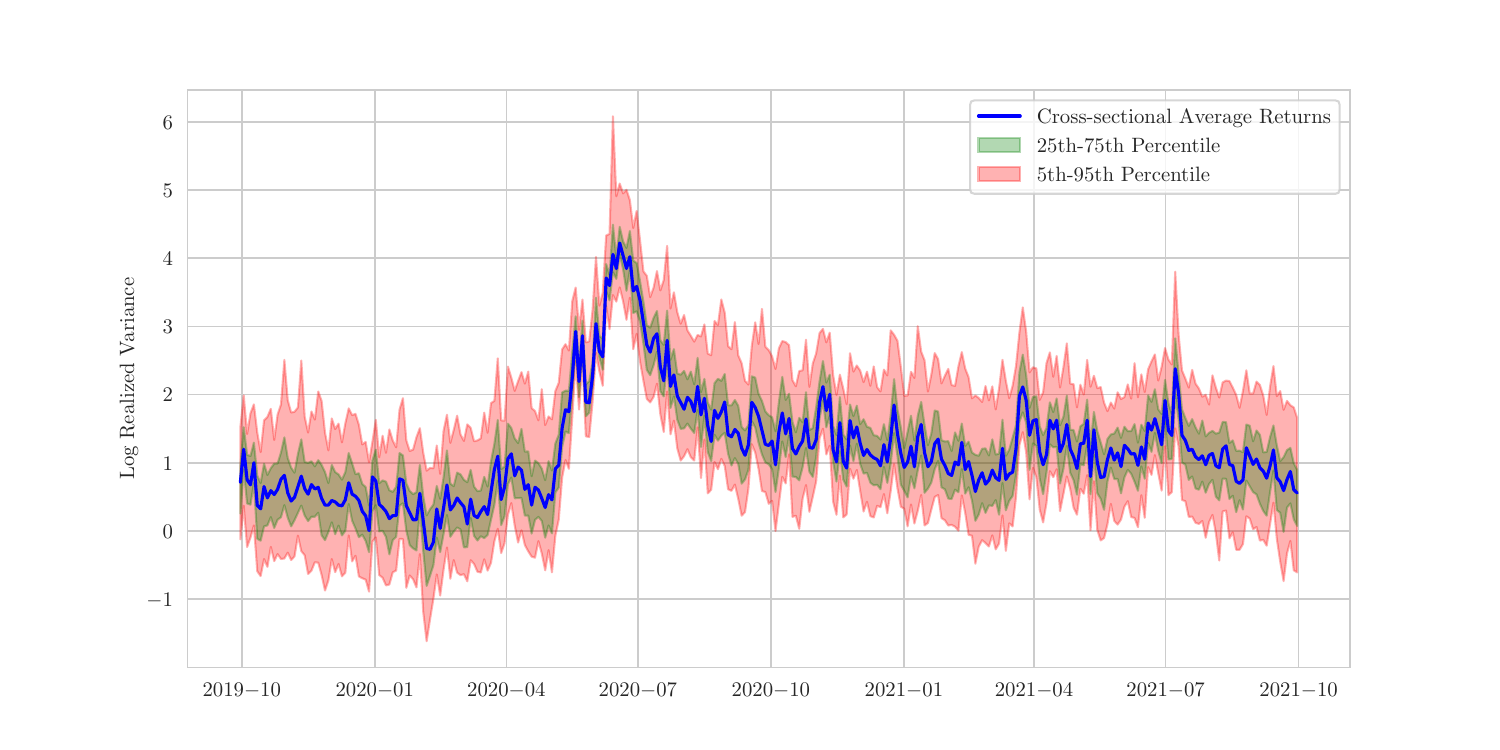}}
		\caption{\textcolor{black}{The blue curve represents cross-sectional average of daily realised volatility across the analysed stocks, with the green area covering the 25-th percentile to the 75-th percentile, and the red area covering the 5-th percentile to the 95-th percentile.}} 
		\label{fig:percentiles_rv_log}
	\end{center}
\end{figure}

	\textcolor{black}{To demonstrate the effectiveness of the proposed architecture, we use high-frequency data from various stocks in the U.S. market for fitting the models. More specifically, DeepVol and the baseline architectures are trained and tested using two years of NASDAQ-100 data. We set the start of our sample at September 30, 2019, with  September 30, 2021 being the last date}. High-frequency data of different sampling frequencies (granularities), i.e., 1, 5, 15, 30, and 60 minutes, is used in our analysis. \textcolor{black}{Tables \ref{table:daily_returns}1 and \ref{table:daily_realised_variance}2 in Appendix \ref{app:a} provide the summary statistics (mininum, maximum, mean, standard deviation, median, skewness, and kurtosis) for the daily close-to-close returns and the realised variance, see Eq. (\ref{eq:rv}). From \textcolor{black}{these} tables, we can see that all of the returns exhibit excess kurtosis, indicating that the \textcolor{black}{distribution} of the returns is fat-tailed. Further, Figure \ref{fig:hists} presents the estimated densities of the daily returns and the daily logarithmic realised volatility, where a departure from normality is evident. Finally, Figures \ref{fig:percentiles_rt} and \ref{fig:percentiles_rv_log} show the cross-sectional average for both returns and realised volatility, along with different percentiles.}\\

DeepVol \textcolor{black}{does} directly perform its prediction from raw high-frequency data, unlike the baseline models, which prior to training require the estimation of daily statistics.
The analyses conducted in this work are based on financial returns, which allow us to transform the original assets' price trend into a quasi-stationary process:
\begin{equation}
	r_{i,t}=\log \left(\frac{p_{i,t}}{p_{i-1, t}}\right), 
	\label{eq:returns}
\end{equation}
where $p_{i,t}$ is the last price of an asset in the $i$-th interval on day $t$, and $r_{i,t}$ is the return over this interval, at the specified sampling frequency,  i.e. 1, 5, 15, 30 or 60 minutes.\\

\textcolor{black}{We have chosen not to include overnight information in our analysis, which aligns with previous studies \citep{engle2012forecasting, zhang2022volatility}. This decision is also supported by recent research indicating a lack of clear methods for integrating overnight session data into daily volatility assessments \citep{pascalau2023increasing}.}\\

\subsubsection{\textcolor{black}{Baselines: Data Preparation} }\label{subsubsec:data_baselines}

The benchmark models used in this work are divided into two categories. Firstly, we consider methods that solely use daily returns to perform day-ahead volatility forecasts. Secondly, we examine methods that take advantage of realised measures in the forecasts.\\

Regarding models depending exclusively on  daily returns, these are obtained  through an analogous procedure to the one followed to retrieve intraday returns through Eq. (\ref{eq:returns}), but using daily returns instead of intraday data.
Concerning methods utilising realised measures, and in consonance with other studies \citep{harvey2018impact, hansen2012realised, shephard2010realising}, we focus on the realised variance for the scope of this work\textcolor{black}{, which is a proxy measure of the volatility that has been shown to be less noisy than daily squared returns \citep{andersen1998answering, barndorff1999non}. It is obtained as follows:}
\begin{equation}
	R V_{t}=\sum_{i=1}^{I} r_{i, t}^2,
	\label{eq:rv}
\end{equation}
where $r_{i, t}$ is the $i$-th intraday return for day $t$, see Eq. (\ref{eq:returns}). \textcolor{black}{Previuous works have shown that the sum of the squared intraday returns is a consistent estimator the unsobervable latent integrated variance of stocks \citep{andersen2001distribution, barndorff2002econometric}}.\\

Various works \citep{andersen2001distribution, ait2005often} study the usage of different sampling frequencies to compute the realised variance through Eq. (\ref{eq:rv}). Selecting a specific intraday's data sampling frequency to compute the realised volatility (e.g., 5 or 30 minutes) involves the optimisation of a trade-off: while we aim to maximise the number of datapoints used, higher sampling frequencies entail an increase of the microstructure noise, which we want to minimise.
We use 5-minutes intraday returns to compute the realised variance through Eq. (\ref{eq:rv}), as this sampling frequency is usually accepted as the optimal value \citep{andersen2001distribution, bandi2006separating}.

\subsubsection{\textcolor{black}{DeepVol: Data Preparation}} \label{subsubsec:data_heavy}


\textcolor{black}{
	As previously discussed, DeepVol differs from classical methods by directly utilizing raw high-frequency data without the need for preprocessing.
	Notably, DeepVol employs a rolling window approach for producing day-ahead volatility forecasts, $\sigma_{t}^{2}$, using a window of past intraday high-frequency returns, $r_{i, t}$, as input data. 
	In Section \ref{sec:experiments}, we explore the optimal window size, referred to hereafter as the receptive field. This parameter determines the number of past days used for predicting day-ahead volatility, essentially indicating how far in the past the first element of the time series that influences the model’s prediction lies.} 
DeepVol is confined to use a specific receptive field, e.g., the previous day’s high-frequency data. This non-recursive ar- chitecture reduces the input data length required by the model, which translates into faster training in comparison to purely autoregressive architectures.
\textcolor{black}{Additionally, we examine the optimal intraday data sampling frequency. Finally, we should remark that DeepVol's operative differs from most state-of-the-art forecasting architectures, which produce predictions at the same granularity (sampling frequency) as the model's input data. Therefore, DeepVol must effectively learn the necessary relationships between high-frequency data and daily volatility, implicitly performing this time-domain transformation.}\\

\section{Baseline Models and Metrics} 
\label{sec:baselines_metrics}
\subsection{Baseline Models}
\label{subsec:baselines}

Most of the models commonly used for volatility forecasting can be traced back to Autoregressive Conditional Heteroscedastic (ARCH) models \citep{engle1982autoregressive}. This family of models assume volatility clustering \citep{cont2007volatility}, i.e., large shocks in prices tend to cluster together. 
ARCH-based models evolved into the well-known Generalized Autoregressive Conditional Heteroscedastic (GARCH) model \citep{bollerslev1986generalized}, which is still widely utilised among industry participants. 
A GARCH$(p,q)$ process is given by:
\begin{equation}
	\sigma_{t}^{2}=\omega+\sum_{i=1}^{p} \alpha_{i} \varepsilon_{t-i}^{2}+\sum_{j=1}^{q} \beta_{j} \sigma_{t-j}^{2},
	\label{eq:garch}
\end{equation}
where $\omega$ is the model's bias, $q$ is the number of lags (order) of the observed volatility, $\sigma_{t}^{2}$; and $p$ is the number of lags of the innovations, $\varepsilon_{t}$. In turn, the returns of prices are related to the innovations by:
\begin{equation}
	r_{t}=\mu+\varepsilon_{t},
\end{equation}
where $\mu$ is the expected return (usually set to zero), and the volatility is related to these innovations by means of the residuals, $e_{t}$:
\begin{equation}
	\varepsilon_{t}=\sigma_{t} e_{t}, \quad e_{t} \sim \mathcal{N}(0,1).
\end{equation}\\

The model's parameters $\{\mu, \omega,\alpha,\beta\}$ can be estimated by performing maximum-likelihood estimation of the joint distribution $f\left(\varepsilon_{1}, \ldots, \varepsilon_{T} ; \{\mu, \omega,\alpha,\beta\}\right)$. The simplest GARCH model consists on a GARCH$(1,1)$ process where $\sigma_{t}=1$ and $\mu=0$.
Several variations leading to new architectures to address the volatility forecasting problem have been developed from the GARCH model. Here, we select some of them for benchmarking purposes.
The integrated GARCH (IGARCH) model \citep{engle1986modelling}, modifies the design of the previous model to grant a longer memory in the autocorrelation of the squared returns, allowing the model to react in a more persistent way to the impact of past squared shocks. Also, IGARCH imposes the following restriction on the model's parameters:
\begin{equation}
	\sum_{i=1}^{p} \alpha_{i} + \sum_{j=1}^{q} \beta_{j}=1,
	\label{eq:igarch_restriction}
\end{equation}
which makes the resulting process a weakly stationary one, since the mean, variance, and autocovariance are finite and constant over time.
The idea behind the IGARCH model motivated the development of the Fractionally Integrated GARCH (FIGARCH) process \citep{baillie1996fractionally}, which is able to capture long-term volatility persistence and clustering features. To do so, it integrates a fractional difference operator (lag operator) $L$ into the conditional variance:
\begin{equation}
	\sigma_{t}^2=\omega+\left[1-\beta L-\phi L(1-L)^{d}\right] \varepsilon_{t}^{2}+\beta \sigma_{t}^2,
\end{equation}
where $0<d<1$ is known as the fractional differencing parameters. The FIGARCH model has been widely used thanks to its ability to capture the volatility's persistence and integrate it into its predictions \citep{cochran2012volatility, biage2019analysis}.
Threshold ARCH (TARCH) models \citep{rabemananjara1993threshold} are also used for benchmarking purposes. The main difference \textcolor{black}{compared} to previous methodologies is that TARCH models divide the distribution of the innovations into disjoint intervals, which are later approximated by a linear function on the conditional standard deviation \citep{zakoian1994threshold}. TARCH models are therefore capable of separately considering the influence of positive and negative innovations:
\begin{equation}
	\sigma_{t}^{2}=\omega+\sum_{i=1}^{p} \alpha_{i} \varepsilon_{t-i}^{2}+\sum_{j=1}^{q} \beta_{j} \varepsilon_{t-j}^{2} \mathbb{I}_{\varepsilon_{t-j}<0}, 
\end{equation}
where $\mathbb{I}_{(\cdot)}$ is the indicator function. The main characteristic of TARCH and other threshold-based approaches, such as TGARCH \citep{park2009persistent}, is their ability to detect abrupt disruptions in the time-series through the indicator function, which may be replaced with a continuous function if a smoother transition is desired.\\

Volatility usually exhibits asymmetric characteristics. This property has led to the development of different asymmetric ARCH type models. For example, the Asymmetric Power ARCH (APARCH) model \citep{ding1993long} assumes a parametric form for the conditional heteroskedasticity's powers. It defines the variance dynamics as follows:
\begin{equation}
	\sigma_{t}^{\delta}=\omega+\sum_{i=1}^{q} \alpha_{i}\left(\left|\varepsilon_{t-i}\right|-\gamma_{i} \varepsilon_{t-i}\right)^{\delta}+\sum_{j=1}^{p} \beta_{j} \sigma_{t-j}^{\delta},
	\label{eq_aparch}
\end{equation}
where we now also have to estimate $\delta > 0$, and $\gamma$. APARCH models nest many other volatility frameworks that can be obtained by imposing restrictions on the APARCH model's parameters.
A similar idea leads to the Asymmetric GARCH model (AGARCH) \citep{engle1993measuring}, which captures the asymmetry in the volatility by using an impact curve associated with the $\alpha_i$ parameter:
\begin{equation}
	\sigma_{t}^{2}=\omega+\sum_{i=1}^{p} \alpha_{i}\left(\varepsilon_{t-i}-\gamma_{i}\right)^{2}+\sum_{j=1}^{q} \beta_{j} \sigma_{t-j}^{2}.
\end{equation}\\

Most of the mentioned models, like IGARCH or APARCH, impose restrictions on the parameters in practice, as Eq. (\ref{eq:igarch_restriction}) states. These restrictions are lifted in the Exponential GARCH  (EGARCH) model \citep{nelson1991conditional}, which is defined as:

\begin{equation}
	\ln \sigma_t^2=\omega+\sum_{i=1}^p \alpha_i\left(\left|\epsilon_{t-i}\right|+\gamma_i \epsilon_{t-i}\right)+\sum_{j=1}^q \beta_j \ln \sigma_{t-j}^2.
	\label{eq:egarch}
\end{equation}

As evidenced by the definition above, the EGARCH model integrates one powerful volatility clustering assumption into its architecture: negative shocks at time $t-1$ produce a stronger impact on the value of the volatility at time $t$ than positive shocks do, allowing for asymmetric effects between positive and negative asset returns. This asymmetry is known in the volatility forecasting literature as leverage effect \citep{bouchaud2001leverage}.\\

All the methods described previously operate through the usage of daily returns. However, as mentioned above, more recent proposals have included the usage of realised measures obtained from the high-frequency data as additional input features for daily volatility forecasting. Among these methodologies, the High-Frequency-Based Volatility (HEAVY) model \citep{shephard2010realising}, has shown superior forecasting capabilities \citep{shephard2010realising, noureldin2012multivariate, sheppard2019factor}. Formally, the model is defined as follows:
\begin{equation}
	\begin{aligned}
		\operatorname{var}\left(r_{t} \mid \mathcal{F}_{t-1}^{\mathrm{HF}}\right) &=\sigma_{t}^{2}=\omega+\alpha \mathrm{RM}_{t-1}+\beta\sigma_{t-1}^{2}, \\[0.1cm]
		\mathbb{E}\left(\mathrm{RM}_{t} \mid \mathcal{F}_{t-1}^{\mathrm{HF}}\right) &=\mu_{t}=\omega_{R}+\alpha_{R} \mathrm{RM}_{t-1}+\beta_{R}\mu_{t-1},
	\end{aligned}
	\label{eq:heavy}
\end{equation}
where $r_t$ denotes daily returns, $RM_t$ denotes  daily realised measures, and $\mathcal{F}_{t-1}^{\mathrm{HF}}$ denotes the high-frequency data utilised to obtain these realised measures. In previous equation, the restrictions $\{\omega, \alpha \geq 0, \beta \in[0,1)\}$ are imposed on the variation of the returns, and $\{\omega_{R}, \alpha_{R}, \beta_{R} \geq 0, \alpha_{R}+\beta_{R} \in[0,1)\}$ on the realised measures' evolution, while observing the following variables:
\begin{equation}
	\begin{array}{l}
		r_{t}=\sqrt{\sigma_{t}^{2}} z_{t}, \\[0.15cm]
		x_{t}=\mu_{t} z_{RV,t}^{2},
	\end{array}
\end{equation}
with:
\begin{equation}
	\left(\begin{array}{c}
		z_{t} \\
		z_{RV, t}
	\end{array}\right) \sim \mathcal{N}(0, I).
	\label{eq:random_variable}
\end{equation}\\

Eq. (\ref{eq:heavy}) shows that HEAVY consists of two parts. While $\sigma_{t}^{2}$ explains the development of the unobserved conditional variance,  $\mu_{t}$ is responsible for explaining the development of the realised measures.
The HEAVY model is clearly motivated by GARCH methodologies, which makes it simple to understand while reporting additional gains in the performance. For further details about it, like parameters inference, we refer the readers to \citep{shephard2010realising}.\\

\textcolor{black}{
	Further, we include both the Realized GARCH \citep{hansen2012realised} and Realized EGARCH \citep{hansen2016exponential} models, which also consider high-frequency/intraday data as predictors to estimate daily volatility. The Realized GARCH model is defined as:
	\begin{equation}
		\begin{aligned}
			\operatorname{var}\left(r_{t} \mid \mathcal{F}_{t-1}^{\mathrm{HF}}\right) &=\sigma_{t}^{2}=\omega+\alpha R M_{t-1}+\beta\sigma_{t-1}^{2}, \\[0.1cm]
			\mathbb{E}\left(\mathrm{RM}_{t} \mid \mathcal{F}_{t-1}^{\mathrm{HF}}\right) &=\xi+\phi \sigma_{t}^{2}+\tau_1 z_t+\tau_2\left(z_t^2-1\right)+u_t,
		\end{aligned}
	\end{equation}
	with $u_t$ being independent and following a standard normal distribution.  Additionally, following  the leverage function proposed in \citet{hansen2012realised}, we include $z_t$ and $z_{t-1}^2$ as a way to incorporate leverage effects, generating an asymmetric response in volatility to return shocks.}\\

\textcolor{black}{
	The main difference of the Realized-GARCH model with respect to the HEAVY model is that the conditional variance at day $t$, that is, $\sigma_{t}^{2}$, is also used as a regressor for estimating the realised measure at time $t$, which entails that both $\alpha$ and $\beta$ parameters are also based on the equation explaining the realised measure evolution. This second equation also depends on the random variable $z_t$, which is not observed directly, as stated in Eq. (\ref{eq:random_variable}).
}\\

\textcolor{black}{
	The Realized EGARCH model, on the other hand, considers the log transformation and integrates the leverage effect into the conditional variance part, which results in:
	\begin{equation}
		\begin{aligned}
			\operatorname{var}\left(r_{t} \mid \mathcal{F}_{t-1}^{\mathrm{HF}}\right) &=\exp \left(\omega+\alpha \log \left(R M_{t-1}\right)+\beta \log \left(h_{t-j}\right)+\tau_1 z_t+\tau_2\left(z_t^2-1\right)\right), \\[0.1cm]
			\log \left(R M_t\right)&=\xi+\phi \log \left(h_t\right)+u_t,
		\end{aligned}
	\end{equation}
	with $u_t$ being normally distributed and independent of $z_t$ when $t \neq s$.
}\\

\textcolor{black}{
	We also include an ARFIMA model \citep{granger1980long, granger1980introduction} as baseline, which is defined as:
	\begin{equation}
		\varphi(L)(1-L)^d\left(R M_t-\mu\right)=\theta(L) \epsilon_t,
	\end{equation}
	where $\varphi(L)=1-\sum_{i=1}^p \varphi_i L^i$ and $\theta(L)=1-\sum_{j=1}^q \theta_j L^j$ are the autoregressive  and moving average lag polynomials. The 
	fractional differencing parameter ${d}$ is responsible of capturing the long-memory properties and $\epsilon_t$ is the error term.
}\\

\textcolor{black}{
	We additionally consider the Heterogeneous Autoregressive model (HAR), originally proposed by \citet{corsi2009simple} as a response to the drawbacks of ARFIMA models. The HAR model has a autoregressive structure and includes averages of daily realised variances over different time horizons. \citet{corsi2009simple} proposes using aggregation periods of one, five, and twenty two for computing the realised variances, see Eq. (\ref{eq:rv}), then define the HAR model as:
	\begin{equation}
		\mathbb{E}\left(\mathrm{RM}_{t} \mid \mathcal{F}_{t-1}^{\mathrm{HF}}\right)=\beta_0+\beta_1 R M_{t-1}+\beta_2 R M_{t-1}^{(w)}+\beta_3 R M_{t-1}^{(m)}+\epsilon_t,
	\end{equation}
	where the parameters $\{\beta_0, \beta_1, \beta_2, \beta_3\}$ are usually estimated with ordinary least squares (OLS) or weighted least squares (WLS) \citep{patton2015good}. Weekly $R M_{t-1}^{(w)}$ and monthly $R M_{t-1}^{(m)}$ realised measures are given by:
	\begin{equation}
		\begin{aligned}
			&R M_t^{(w)}=\frac{1}{5}\left(R M_{t-1}^{(d)}+R M_{t-2}^{(d)}+R M_{t-3}^{(d)}+R M_{t-4}^{(d)}+R M_{t-5}^{(d)}\right)\\
			&R M_t^{(m)}=\frac{1}{22}\left(R M_{t-1}^{(d)}+R M_{t-2}^{(d)}+\ldots+R M_{t-21}^{(d)}+R M_{t-22}^{(d)}\right)
		\end{aligned}
		\label{eq:har_rm}
	\end{equation}
}\\

\textcolor{black}{
	Finally, we include as baselines a MLP \citep{rumelhart1986learning} and a LSTM \citep{hochreiter1997long} that operate directly over the high-frequency data, as done in previous research papers \citep{zhang2022volatility, masini2023machine}. 
}

\subsection{Evaluation Metrics} \label{subsec:metrics}

In this section, we define a series of metrics that will be used to assess the day-ahead volatility forecast of our proposed architecture against previously defined baseline models. The Mean Absolute Error (MAE) and the Root Mean Squared Error (RMSE) constitute two of the most common error functions to evaluate the performance of volatility forecasting architectures. While a number of articles focus entirely on those two metrics to report performance \citep{shen2021bitcoin, izzeldin2019forecasting}, we complement them with the usage of the Symmetric Mean Absolute Percentage Error (SMAPE). This relative error measure has both a lower and upper bound, contrary to the Mean Absolute Percentage Error (MAPE), and it is scale independent.
Also, the Maximum Error (ME) is used to illustrate which models produce the more significant inaccuracies: poor performance adapting to new regimes, as volatility shocks, lead certain models to substantial momentary discrepancies between the forecast and the actual volatility, which leads to an increase in the ME. We complement the ME with the Median Absolute Error (MedAE), an outliers-robust metric.
Lastly, we include the Quasi Log-Likelihood (QLIKE), which has proven to be a noise robust loss function in the volatility proxy. Both the QLIKE and the RMSE will be used as loss functions to optimise the model's parameters during Section \ref{sec:experiments}, while the rest of the metrics will be used to assess the models' performance. We summarise the definitions of all metrics below.


\begingroup
\allowdisplaybreaks
\color{black}
\begin{align*}
	&\ell_{\operatorname{\it{MAE}}} ({\sigma}_{t}^{2},\hat{\sigma}_{t}^{2}) =\frac{1}{T} \sum_{t=1}^{T}\left|{\sigma}_{t}^{2}-\hat{\sigma}_{t}^{2}\right| \vphantom{\sqrt{\frac{1}{T} \sum_{t=1}^{T}}},  \\
	&\ell_{\operatorname{\it{rmse}}} ({\sigma}_{t}^{2},\hat{\sigma}_{t}^{2}) =\sqrt{ \frac{1}{T} \sum_{t=1}^{T}\left({\sigma}_{t}^{2}-\hat{\sigma}_{t}^{2}\right)^2 } \vphantom{\frac{1}{T} \sum_{t=1}^{T}},\\
	&\ell_{\operatorname{\it{SMAPE}}} ({\sigma}_{t}^{2},\hat{\sigma}_{t}^{2})=\frac{1}{T} \sum_{t=1}^{T}\frac{\left|{\sigma}_{t}^{2}-\hat{\sigma}_{t}^{2}\right|}{({\sigma}_{t}^{2}+\hat{\sigma}_{t}^{2})/2} \vphantom{\sqrt{\frac{1}{T} \sum_{t=1}^{T}}},  \\
	&\ell_{\operatorname{\it{ME}}} ({\sigma}_{t}^{2},\hat{\sigma}_{t}^{2})= \text{\it{max}}\left(\left|{\sigma}_{t}^{2}-\hat{\sigma}_{t}^{2}\right|\right) \vphantom{\sqrt{\frac{1}{T} \sum_{t=1}^{T}}}, \\
	&\ell_{\operatorname{\it{MedAE}}} ({\sigma}_{t}^{2},\hat{\sigma}_{t}^{2})=\text{\it{median}}(\left|{\sigma}_{t}^{2}-\hat{\sigma}_{t}^{2}\right|) \vphantom{\sqrt{\frac{1}{T} \sum_{t=1}^{T}}},  \\ 
	&\ell_{\operatorname{\it{QLIKE}}} ({\sigma}_{t}^{2},\hat{\sigma}_{t}^{2}) =\frac{1}{T} \sum_{t=1}^{T} \log \left(\hat{\sigma}_{t}^{2}\right)+\frac{\sigma_{t}^{2}}{\hat{\sigma}_{t}^{2}} \vphantom{\sqrt{\frac{1}{T} \sum_{t=1}^{T}}},
\end{align*}
\endgroup
where $\hat{\sigma}_{t}^{2}$ and ${\sigma}_{t}^{2}$ represent the volatility forecast and the volatility proxy measure, respectively, with $T$ the total amount of rolling forecasts.

\subsection{\textcolor{black}{Model Confidence Set}} \label{subsec:mcs}


\textcolor{black}{
	When assessing the predictions of various models for a specific task, a lower value of the loss function for one model in comparison to another implies superior forecasting performance of the former over the latter. However, drawing such conclusions based on a single loss function or a single sample may not be reliable. 
	Therefore, in addition to employing multiple loss functions in our evaluation, as mentioned in Section \ref{subsec:baselines}, we also utilise statistical tests to determine whether a particular model outperforms another.
}\\

\textcolor{black}{
	Various works have focused on developing frameworks to determine whether a particular model outperforms another. Among them, the Diebold-Mariano (DM) test \citep{diebold2002comparing} and the Superior Predictive Ability (SPA) test \citep{hansen2005test} are popular and reliable approaches. However, the Model Confidence Set (MCS) method \citep{hansen2011model} has several advantages over them. For instance, the MCS method has fewer assumptions and does not require the selection of a benchmark, making it suitable for scenarios where identifying a clear baseline is challenging. Additionally, it allows for multiple optimal models, which contrasts with the DM test, which is only applicable for pairwise comparisons between two models. In our study, where we evaluate 16 models in parallel (see Section \ref{sec:experiments}), the MCS test is more appropriate. Therefore, we adopt the MCS test in our empirical evaluation, following the approach of \citet{hansen2011model} and \citet{zhang2022volatility}, among others.}\\

\textcolor{black}{
	The MCS test aims to identify a subset of models $\mathcal{M}^*$ with significantly superior performance from a set of model candidates $\mathcal{M}_0$, at a specified level of confidence. It operates through an elimination process that is based on sequentially testing the null hypothesis:
	\begin{equation}
		H_{0,\mathcal{M}} : \mathbb{E}(\Delta L_{ij,t}) = 0, \quad \text{for all } i, j \in \mathcal{M} \subset \mathcal{M}_0,
	\end{equation}
	where $\Delta L_{ij,t}$ represents the difference in loss between models $i$ and $j$ at time $t$, as measured by a specific loss function $L$, see Section \ref{subsec:metrics}.} \\

\textcolor{black}{The MCS procedure involves a series of significance tests aimed at identifying objects significantly inferior to others in $\mathcal{M}_0$ for elimination. The set of superior objects is defined by:
	\begin{equation}
		\mathcal{M}^* = \{ i \in \mathcal{M}_0 :\mathbb{E}(\Delta L_{ij,t}) \leq 0 \text{ for all } j \in \mathcal{M}_0 \}.\\
\end{equation}}

\textcolor{black}{
	Through an equivalence test, $\delta_\mathcal{M}$, and elimination rule, $e_\mathcal{M}$, we can test the hypothesis $H_0$ for any $\mathcal{M} \subset \mathcal{M}_0$, identifying the model that is to be removed from out set of candidates in case $H_0$ is rejected. After repeating the two tests, we obtain a set of ``surviving'' objects; the model confidence set. For further details, we refer the readers to \citet{hansen2011model}.}\\

\section{Model} \label{sec:model}

\subsection{Problem Definition}\label{subsec:problem_definition}

Considering a set of assets, 
$\Delta \in \mathbb{R}^{d}$, where $d \in \mathbb{N}$ denotes the dimension of the input vector, with $T \in  \mathbb{N}$ days' intraday high-frequency data associated to them, $\{\mathbf{r}_{t}^{1:J}\}_{t=1}^{T} $, where $\mathbf{r}_{t}^{1: J}=(r_{t}^{1}, r_{t}^{2}, \ldots, r_{t}^{J})$ are the intraday returns of the $t$-th day, with $T$ being referred to as receptive field, and with ${J \in  \mathbb{N}} $ the length of each day's intraday data, our goal is to forecast the day-ahead realised volatility: 
\begin{equation}
	\hat{\sigma}^{2}_{T+1} = f_{\theta}\left( r_{t=1}^{1}, r_{t=1}^{2}, \ldots, r_{t=1}^{J}, r_{t=2}^{1}, \ldots, r_{t=T}^{1}, \ldots, r_{t=T}^{J} \right),
\end{equation}

where  
$f_{\theta} :\mathbb{R}^{d} \rightarrow \mathbb{R}^{m}, m \in \mathbb{N},$ is a function implemented through a Dilated Causal Convolutions (DCC)-based neural network, with $\theta \in \Theta$ being the learnable parameters of the model from a set $\Theta \in \mathbb{R}^n$, for some $n \in \mathbb{N}$. \\

These parameters fully specify the corresponding volatility forecast. Therefore, we aim to obtain the set of optimal parameters $\hat{\theta} \in \Theta$ that minimises the difference between the forecasted volatility, $\hat{\sigma}_t^2$, and the volatility's proxy measure ${\sigma}_t^2$ for the considered assets:
\begin{equation}
	\hat{\theta}=\underset{\theta \in \Theta}{\operatorname{argmin}} \, \mathcal{L} \left(f_\theta(\Delta), {\sigma}_t^2(\Delta)\right),
\end{equation}
where $\mathcal{L}$ is the selected metric for evaluating the forecast accuracy.

\subsection{Dilated Causal Convolutions}\label{subsec:dcc}

Our volatility forecasting proposal, DeepVol, uses DCCs as a technique to integrate the high-frequency information into the realised volatility prediction.
The deployment of such an architecture allows the usage of a large receptive field
\textcolor{black}{that can handle lengthy sequences of data effectively} \textcolor{black}{without a significant increase in the number of parameters} \textcolor{black}{thanks to the use of dilated connections,} yielding improved computational efficiency. 

The proposed architecture consists of $L$ convolutional layers. The convolution operation performed by the first layer, between the input sequences $x$ and the kernel $k$, can be defined as follows:
\begin{equation}
	F^{(l=1)}(t)=\left(x *_{d} k^{(l=1)}\right)(t)=\sum_{\tau=0}^{s-1} k^{(l=1)}_{\tau} \cdot x_{t-d \tau},
\end{equation}
being $d$ the dilation factor and $k$ the filter, with size $s \in\mathbb{Z}$.
For each of the rest $l$-th layers, we can define the convolution operation as:
\begin{equation}
	F^{(l)}(t)=\left(F^{(l-1)} *_{d} k^{(l)}\right)(t)=\sum_{\tau=0}^{s-1} k^{(l)}_{\tau} \cdot F_{t-d \tau}^{l-1}(t).
\end{equation}\\

As previous equations state, the inner product performed by the dilated causal convolutions is based on entries that are  a fixed number of steps apart from each other, contrary to CNN and Causal-CNN, which operate with consecutive entries.
Furthermore, each of the layers in this hierarchical structure defines the kernel operation as an affine function acting between layers:
\begin{equation}
	k^{(l)}: \mathbb{R}^{N_{l}} \longrightarrow \mathbb{R}^{N_{l+1}}, 1 \leq l \leq L.
	\label{eq:affine}
\end{equation}\\

\begin{figure}[] 
	\begin{center}
		\centerline{\includegraphics[width=0.9\columnwidth]{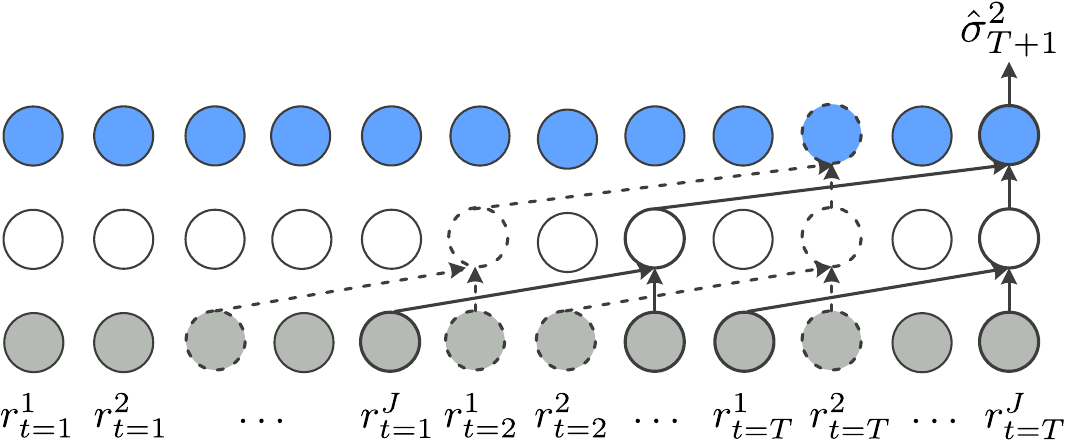}}
		\caption{\textcolor{black}{DeepVol's operation over the high-frequency data through dilated causal convolutions}. The dilation factor grows exponentially, allowing an increase in the receptive field without increasing the model's complexity.} 
		\label{fig:architecture}
	\end{center}
\end{figure}

\textcolor{black}{We should mention that, through the introduction of residual connections, initially proposed by \citet{he2016deep}, certain layers of the model can deviate from this traditional sequential flow of information. In standard sequential architectures, the output of layer $l$ typically serves as the input for layer $l+1$, as stated in Eq. (\ref{eq:affine}). However, residual connections introduce an exception by connecting layers that are not directly adjacent, thereby creating parallel paths for information flow. This bypasses some intermediate layers and allows for the direct transmission of information between non-adjacent layers. By introducing additional paths for information flow, residual connections enable the usage of deeper models with larger receptive fields and help alleviate challenges associated with training deep networks, such as the vanishing gradient problem \citep{hochreiter2001gradient}.}
The complete operative of the proposed model can be defined as follows:
\begin{equation}
	\sigma^{2}_{T+1}(\{\mathbf{r}_{t}^{1:J}\}_{t=1}^{T})=\alpha_{0}+\sum_{l=1}^{L} \alpha_{l} \, \sigma_{ReLu} (F^{(l)}(\{\mathbf{r}_{t}^{1:J}\}_{t=1}^{T})),
\end{equation}
where $\sigma_{ReLu}: \mathbb{R} \mapsto \mathbb{R}$ is the selected non-linearity and $\{\alpha_0, \cdots, \alpha_l, \cdots, \alpha_L\}$ is a set of weights applied to the convolutional operations. \textcolor{black}{Further, the attention mechanism \textcolor{black}{---we use the classic attention mechanism \citep{bahdanau2014neural}, which contrast with self-attention \citep{vaswani2017attention}, as used in Transformer models, where the computational cost scales quadratically in the length of the input sequence, see \citet{moreno2024rough_short, moreno2024rough}---} weights each layer's inputs to reflect their importance at each time-step.}
\textcolor{black}{Figure \ref{fig:architecture} illustrates how the dilated causal convolutions used by the model operate, where a receptive field with previous $T$ days' intraday data is processed through a hierarchy of these dilated convolutions to forecast the day-ahead realised volatility.}

DeepVol, like any deep-feed-forward neural network, is approximating the volatility's unknown function through sample pairs of input and output data $(x, y)$. Formally speaking, DeepVol is approximating some function $f_{\theta}(\cdot)$ which is not available in closed form by finding the optimal model's parameters $\hat{\theta}$ derived from the best function approximation $f^*_{\theta}(\cdot)$.

\section{Experiments} \label{sec:experiments}

In this section, DeepVol's volatility forecasts will be evaluated and compared with the baseline models described in Section \ref{subsec:baselines}. For benchmarking purposes, we will utilise the metrics described in Section \ref{subsec:metrics}. Besides classic out-of-sample forecast comparisons, we perform different experiments to present some additional insights into the inner workings of DeepVol.
We \textcolor{black}{validate} DeepVol's behaviour while varying the intraday data sampling frequency, studying the discrepancy in model behaviour when trained on different granularity regimes. In close relation to this, we also explore the usage of different receptive field sizes and how this affects the model performance.
\textcolor{black}{Further}, we analyse the inclusion of realised measures as an extra input to the model, studying if its addition can improve the forecasting accuracy. \textcolor{black}{Finally, we analyse the generalization capabilities of our model on out-of-distribution-stocks.}
Besides the models presented in Section \ref{subsec:baselines}, we also include a martingale process for comparison purposes.

\subsection{\textcolor{black}{Training Details}}
\label{subsec:setup}

We use the Quasi Log-Likelihood as loss function to train the model parameters. We choose Adaptive Moment Estimation Algorithm (ADAM) \citep{kingma2014adam} as optimiser, even though different experiments were conducted exploring the usage of Averaged Stochastic Gradient Descent (ASGD) \citep{kingma2014adam} and Limited Memory BFGS (L-BFGS) \citep{liu1989limited}. While the usage of these optimisers usually entails smoother predictions, the reported performance declined with respect to ADAM, hence, they were not considered further. Early stopping is used during the training process \textcolor{black}{through the validation set}. DeepVol is implemented in Pytorch-Lightning \citep{Falcon_PyTorch_Lightning_2019}, and the experiments are conducted using a NVIDIA Titan Xp GPU.\\


\textcolor{black}{
	Through the use of validation sets ---we detail how the data is divided into train/validation/test sets in each experiment---, we assess a range of hyperparameters to identify the ones that perform best on the validation data. These hyperparameters, together with their optimal value for the out-of-sample forecast experiment, are summarized in Table \ref{tab:hyperparameters}\textcolor{black}{, we refer the readers to \citet{kingma2014adam, goodfellow2016deep, prechelt2002early, oord2016wavenet, he2016deep, mikolov2013efficient, bahdanau2014neural, vaswani2017attention} for further details on the employed methodologies and the validation of their hyperparameters.} Further, in Section \ref{subsec:cr}, we present a detailed analysis of this validation process, focusing specifically on different sampling frequencies and receptive field sizes, having selected previously the optimal values for the remaining hyperparameters. We believe that examining these two parameters offers an insightful comparison with previous studies that investigated the same parameters while computing realised measures from high-frequency data \citep{andersen2001distribution, bandi2006separating}. Finally, we also validate the number of hidden layers and neurons per layer for the MLP, and the number of layers and units for the LSTM.
}\\


\begin{table}[t!]
	\fontsize{10.0}{16}\selectfont
	\centering
	\centering
	\caption{\textcolor{black}{DeepVol's hyperparameters for out-of-sample forcasting.}}
	\label{tab:hyperparameters}
	{\color{black}
		\begin{tabular}{@{}cc@{}}
			\thickhline
			\textbf{Hyperparameter} & \textbf{Optimal Value} \\ \thickhline
			Learning Rate & 1e-3 \\
			Batch Size & 512 \\
			Number of Epochs & 1000 \\
			Early Stopping (Patience) & 50 \\
			Optimizer & Adam \\
			Kernel Size & 3 \\
			Conditioning Range & 1 \\
			Sampling Frequency & 5 mins \\
			Num. Blocks & 2 \\
			Num. Layers & 6 \\
			Residual Channels & 32 \\
			Dilation Channels & 64 \\
			Skip Channels & 128 \\
			End Channels & 64 \\
			Loss Function & QLIKE \\
			\thickhline
	\end{tabular}}
\end{table}

\subsection{\textcolor{black}{Receptive Field and Sampling Frequency Analysis}}
\label{subsec:cr}

\textcolor{black}{Besides being two hyperparameters in our model, }the receptive field size and the intraday sampling frequency \textcolor{black}{can} shed light on the inner workings of DeepVol when analysed further.
As mentioned in Section \ref{subsubsec:data_baselines}, a number of studies have validated the optimal intraday data sampling frequency for computation of the realised measures from high-frequency data \citep{andersen2006market, hansen2006realised, corradi2006semi}, commonly concluding that using a granularity of 5 or 10 minutes minimises the microstructure noise effect while maximising the usage of high-frequency information.\\

\textcolor{black}{In this section, we study this same trade-off in the proposed deep-learning architecture, analysing the effect that using different sampling frequencies and receptive field sizes has on the model performance. To do so, we use the NASDAQ-100 dataset described in Section \ref{subsec:data}, splitting it into three main folds. The training set contains 12 months of intraday data, from September 30, 2019, to September 30, 2020. The validation set contains three months of data, from October 01, 2020, to December 31, 2020. Finally, the test set contains nine months of data, from January 1, 2021, to September 30, 2021. Metrics introduced in Section \ref{sec:baselines_metrics} are used to analyse the models' performance.}\\



\begin{table}[t!]
	\fontsize{10.0}{16}\selectfont
	\centering
	\caption{Receptive field and sampling frequency study.}
	\vspace*{3mm}
	\label{table:cr}
	\begin{tabular}{cclcccccc}
		\thickhline
		\begin{tabular}[c]{@{}c@{}}\textbf{Sampling}\\ \textbf{Frequency}\end{tabular} & \begin{tabular}[c]{@{}c@{}}\textbf{Receptive}\\ \textbf{Field}\end{tabular} & \textbf{} & \textbf{MAE} & \textbf{RMSE} & \textbf{SMAPE} & \textbf{QLIKE} & \textbf{ME} & \textbf{MedAE} \\ \thickhline
		\textbf{1 min}                   & \textbf{1}           &           & 4.096      & 8.462       & 0.287        & 342.313        & 71.749     & 2.396       \\ \hline
		\multirow{3}{*}{\textbf{5 min}}  & \textbf{1} &           & \textbf{3.903}        & 8.457         & \textbf{0.279}          & \textbf{340.779}        & 71.779      & 2.008         \\
		& \textbf{2}           &           & 4.429      & 9.495       & 0.308        & 367.209        & \textbf{70.036}     & 1.756       \\
		& \textbf{3}           &           & 4.054        & \textbf{8.379}         & 0.285          & 343.359        & 70.457      & 2.334         \\ \hline
		\multirow{4}{*}{\textbf{15 min}}    & \textbf{1}           &           & 3.993        & 8.436         & 0.283          & 343.412        & 70.915      & 2.185         \\
		& \textbf{2}           &           & 4.651        & 10.437        & 0.312          & 365.893        & 70.926      & 1.836         \\
		& \textbf{3}           &           & 5.817        & 9.520         & 0.323          & 362.235        & 72.338      & 4.577         \\
		& \textbf{5}           &           & 6.736        & 10.217        & 0.336          & 366.192        & 72.240      & 5.594         \\ \hline
		\multirow{5}{*}{\textbf{30 min}} & \textbf{1}           &           & 4.259        & 9.699         & 0.318          & 689.633        & 75.793      & 1.632         \\
		& \textbf{2}           &           & 4.503        & 10.140        & 0.327          & 789.326        & 79.345      & 1.724         \\
		& \textbf{3}           &           & 4.473        & 9.931         & 0.324          & 784.843        & 75.059      & 1.705         \\
		& \textbf{5}           &           & 4.705        & 10.802        & 0.326          & 833.966        & 83.416      & 1.676         \\
		& \textbf{10}          &           & 4.732        & 11.084        & 0.327          & 853.981        & 85.656      & \textbf{1.591}         \\ \hline
		\multirow{5}{*}{\textbf{60 min}} & \textbf{1}           &           & 4.988        & 10.516        & 0.297          & 366.509        & 82.207      & 2.402         \\
		& \textbf{2}           &           & 5.616        & 12.596        & 0.324          & 709.082        & 99.828      & 2.178         \\
		& \textbf{3}           &           & 5.441        & 12.615        & 0.319          & 688.996        & 99.612      & 2.017         \\
		& \textbf{5}           &           & 5.520        & 12.456        & 0.326          & 714.871        & 95.763      & 2.071         \\
		& \textbf{10}          &           & 4.997        & 11.186        & 0.322          & 706.917        & 87.096      & 1.927         \\ \thickhline
	\end{tabular}
\end{table}

Table \ref{table:cr} collects the results of an analysis whose main objective is \textcolor{black}{not only to select the hyperparameters values with the best performance on the validation set, but also} to evaluate if DeepVol's performance is robust to increasing the receptive field or modifying the sampling frequency\textcolor{black}{. Firstly,} it is interesting to \textcolor{black}{remark} that using intraday data from one day with a sampling frequency of 5-minutes proves to be optimal. This scenario reports the best results with regard to all the considered metrics but the MedAE.
\textcolor{black}{Further, this result is particularly relevant considering that it aligns with previous literature suggesting an optimal granularity around 5 minutes, like \citet{andersen2001distribution, bandi2006separating}, or \cite{andersen1997heterogeneous}, where it is affirmed 
	that ``the five-minute horizon is short enough that the accuracy of the continuous record asymptotics underlying our realised volatility measures work well, and long enough that the confounding influences from market microstructure frictions are not overwhelming''.}
\textcolor{black}{
	Even so, it's important to note that our observation of 5-minute intervals appearing optimal may be influenced by our task of predicting realised variance computed from intraday data sampled at 5-minute intervals. Therefore, we want to clarify that we do not propose a 5-minute frequency as universally optimal for volatility forecasting. Instead, in the specific context of predicting realised volatility in our study, a 5-minute granularity proves effective. In real-world scenarios, practitioners should properly validate both the sampling frequency and the conditioning range, taking into account the specific volatility proxy being used.
}\\

Secondly, the increment of the receptive field leads to a degradation of performance. This indicates that, for the proposed architecture \textcolor{black}{and the selected proxy measure of volatility}, all the relevant information for forecasting the day-ahead volatility can be obtained from the previous day high-frequency data \textcolor{black}{ and increasing the receptive field yields more conservative predictions that} deteriorate its performance. \\

Thirdly, the best performance in terms of MedAE is obtained when using a 30 minutes sampling frequency, together with a receptive field of ten days. This result can be directly related to the hypothesis previously mentioned: \textcolor{black}{a} longer receptive field leads to a more conservative forecast, resulting in a lower Median Absolute Error. In this scenario, the model is less prone to forecast volatility jumps, a behaviour commonly associated with integrating momentum indicators\textcolor{black}{; however,} this specific setup leads to a deterioration in all the other metrics. The growth in the receptive field size prevents the model from forecasting more drastic changes in the presence of volatility shocks, leading to more conservative predictions \textcolor{black}{compared to those} reported when using just the previous day's intraday data.\\

\subsection{Out-of-sample Forecast}
\label{subsec:nasdaq}

This section aims to provide an out-of-sample performance comparison between the proposed model and some classical methodologies widely used in the finance industry. \textcolor{black}{As mentioned in previous section, the NASDAQ-100 dataset is spitted into three main folds. The results presented in this section are computed over the test set, containing nine months of data, from January 1, 2021, to September 30, 2021. The validation set is used both to identify the optimal hyperparameters and to monitor the early stopping criteria.}
For this specific study, a sampling frequency of 5-minutes for the intraday data and a receptive field of one day (using $t$-day intraday's data to predict the realised volatility for day $t+1$) are utilised, as \textcolor{black}{ they are found to be optimal in Section \ref{subsec:cr}}. \textcolor{black}{Additionally, Table \ref{tab:hyperparameters} provides a summary of the remaining hyperparameters' values.}\\

\begin{table}[]
	\fontsize{10.0}{15}\selectfont
	\centering
	\caption{Out-of-sample forecast: experiments results for the NASDAQ-100 dataset. An asterisk ($^*$) indicates models that are included in the MCS at the 5\% significance level.}
	\vspace*{3mm}
	\label{table:baselines}
	\begin{tabular}{clcccccc}
		\thickhline
		\textbf{Method}     & \textbf{} & \textbf{MAE}   & \textbf{RMSE}  & \textbf{SMAPE} & \textbf{QLIKE}   & \textbf{ME}     & \textbf{MedAE} \\ \thickhline
		\textbf{martingale} &           & 5.180          & 11.410         & 0.324          & 747.480          & 96.654         & \textbf{1.614}$^*$ \\
		\textbf{TARCH}      &           & 4.849          & 10.320         & 0.301          & 351.310          & 71.659         & 2.804          \\
		\textbf{IGARCH}     &           & 5.008          & 10.534         & 0.302          & 351.702          & 72.048         & 2.797          \\
		\textbf{FIGARCH}    &           & 4.631          & 10.356         & 0.294          & 349.245          & 71.050         & 2.460          \\
		\textbf{APARCH}     &           & 4.730          & 10.096         & 0.299          & 349.974          & \textbf{70.088}$^*$         & 2.859          \\
		\textbf{AGARCH}     &           & 4.833          & 10.304         & 0.324          & 351.217          & 71.215         & 2.819          \\
		\textbf{EGARCH}     &           & 4.793          & 10.180         & 0.300          & 348.614          & 70.615$^*$         & 2.917          \\
		\textbf{HEAVY}      &           & 4.565          & 10.239         & 0.292          & 343.490          & 72.404         & 2.545          \\
		\textcolor{black}{\textbf{Real-GARCH}}      &           & \textcolor{black}{4.598}          & \textcolor{black}{10.062}         & \textcolor{black}{0.299}          & \textcolor{black}{344.282}          & \textcolor{black}{70.896}         & \textcolor{black}{2.488}    \\
		\textcolor{black}{\textbf{Real-EGARCH}}    &           & \textcolor{black}{4.686}          & \textcolor{black}{10.103}         & \textcolor{black}{0.299}          & \textcolor{black}{345.739}          & \textcolor{black}{71.061}         & \textcolor{black}{2.512}    \\
		\textcolor{black}{\textbf{ARFIMA}} & & \textcolor{black}{4.620} & \textcolor{black}{10.345} & \textcolor{black}{0.296} & \textcolor{black}{344.800} & \textcolor{black}{71.901} & \textcolor{black}{2.480} \\
		\textcolor{black}{\textbf{HAR}} & & \textcolor{black}{4.587} & \textcolor{black}{10.306} & \textcolor{black}{0.293} & \textcolor{black}{343.652} & \textcolor{black}{72.116} & \textcolor{black}{2.451} \\
		\textcolor{black}{\textbf{MLP}} & & \textcolor{black}{4.820} & \textcolor{black}{10.607} & \textcolor{black}{0.307} & \textcolor{black}{350.994} & \textcolor{black}{73.302} & \textcolor{black}{2.939} \\
		\textcolor{black}{\textbf{LSTM}} & & \textcolor{black}{4.441} & \textcolor{black}{10.032} & \textcolor{black}{0.295} & \textcolor{black}{345.465} & \textcolor{black}{72.650} & \textcolor{black}{2.570} \\
		\textcolor{black}{\textbf{DeepVol-RM}} & & \textcolor{black}{4.383} & \textcolor{black}{9.823} & \textcolor{black}{0.284} & \textcolor{black}{342.433}$^*$ & \textcolor{black}{72.016} & \textcolor{black}{2.405}\\ 
		\textbf{DeepVol}    &           & \textbf{3.903}$^*$ & \textbf{8.457}$^*$ & \textbf{0.279}$^*$ & \textbf{340.779}$^*$ & 71.779 & 2.008          \\ \thickhline
	\end{tabular}
\end{table}

\begin{table}[t!]
	\fontsize{10.0}{15}\selectfont
	\centering
	\caption{Out-of-sample forecast: percentage of improvement/degradation over the  martingale process and the HEAVY model, for each of the evaluated models.}
	\vspace*{3mm}
	\label{table:improvement}
	\begin{tabular}{clcccccc}
		\thickhline
		\textbf{Method}     &  & \textbf{MAE} & \textbf{RMSE} & \textbf{SMAPE} & \textbf{QLIKE} & \textbf{ME} & \textbf{MedAE} \\ \thickhline
		\textbf{}           &  & \multicolumn{6}{c}{\textbf{Improvement over martingale (\%)}}                                \\ \cline{3-8} 
		\textbf{martingale} &  & -            & -             & -              & -              & -           & -             \\
		\textbf{TARCH}      &  & 6.398        & 9.553         & 7.099          & 53.001         & 25.860       & -73.730       \\
		\textbf{IGARCH}     &  & 3.320        & 7.677         & 6.790          & 52.948         & 25.458       & -73.296       \\
		\textbf{FIGARCH}    &  & 10.598       & 9.238         & 9.259          & 53.277         & 26.490       & -52.410       \\
		\textbf{APARCH}     &  & 8.687        & 11.516        & 7.716          & 53.179         & \textbf{27.486}      & -77.138       \\
		\textbf{AGARCH}     &  & 6.699        & 9.693         & 0.000          & 53.013         & 26.319       & -74.659       \\
		\textbf{EGARCH}     &  & 7.471        & 10.780        & 7.407          & 53.361         & 26.940      & -80.731       \\
		\textbf{HEAVY}      &  & 11.873       & 10.263        & 9.877          & 54.047         & 25.089      & -57.677       \\
		\textcolor{black}{\textbf{Real-GARCH}} & & \textcolor{black}{11.236} & \textcolor{black}{11.814} & \textcolor{black}{7.716} & \textcolor{black}{53.941} & \textcolor{black}{26.650} & \textcolor{black}{-51.673} \\ 
		\textcolor{black}{\textbf{Real-EGARCH}} & & \textcolor{black}{9.537} & \textcolor{black}{11.455} & \textcolor{black}{7.716} & \textcolor{black}{53.746} & \textcolor{black}{26.479} & \textcolor{black}{-55.638} \\ 
		\textcolor{black}{\textbf{ARFIMA}} & & \textcolor{black}{10.811} & \textcolor{black}{9.334} & \textcolor{black}{8.642} & \textcolor{black}{53.872} & \textcolor{black}{25.610} & \textcolor{black}{-53.656} \\ 
		\textcolor{black}{\textbf{HAR}} & & \textcolor{black}{11.448} & \textcolor{black}{9.676} & \textcolor{black}{9.568} & \textcolor{black}{54.025} & \textcolor{black}{25.387} & \textcolor{black}{-51.859} \\ 
		\textcolor{black}{\textbf{MLP}} & & \textcolor{black}{6.950} & \textcolor{black}{7.038} & \textcolor{black}{5.247
		} & \textcolor{black}{53.043} & \textcolor{black}{24.160} & \textcolor{black}{-82.094} \\ 
		\textcolor{black}{\textbf{LSTM}} & & \textcolor{black}{14.266} & \textcolor{black}{12.077} & \textcolor{black}{10.494} & \textcolor{black}{53.900} & \textcolor{black}{25.107} & \textcolor{black}{-60.408}\\ 
		\textcolor{black}{\textbf{DeepVol-RV}} & & \textcolor{black}{15.386} & \textcolor{black}{13.908} & \textcolor{black}{12.346} & \textcolor{black}{54.188} & \textcolor{black}{25.491} & \textcolor{black}{-49.009}\\ 
		\textbf{DeepVol}    &  & \textbf{24.653}       & \textbf{25.881}        & \textbf{13.889}         & \textbf{54.410}         & 25.736      & \textbf{-24.411}       \\ \thickhline
		\textbf{}           &  & \multicolumn{6}{c}{\textbf{Improvement over HEAVY (\%)}}                                     \\ \cline{3-8} 
		\textbf{martingale} &  & -13.472      & -11.437       & -10.959        & -117.613       & -33.492     & \textbf{36.579}        \\
		\textbf{TARCH}      &  & -6.212       & -0.791        & -3.082         & -2.277         & 1.029     & -10.181       \\
		\textbf{IGARCH}     &  & -9.704       & -2.881        & -3.425         & -2.391         & 0.492     & -9.906        \\
		\textbf{FIGARCH}    &  & -1.446       & -1.143        & -0.685         & -1.675         & 1.870     & 3.340         \\
		\textbf{APARCH}     &  & -3.614       & 1.397         & -2.397         & -1.888         & \textbf{3.120}     & -12.342       \\
		\textbf{AGARCH}     &  & -5.871       & -0.635        & -10.959        & -2.250         & 1.642     & -10.771       \\
		\textbf{EGARCH}     &  & -4.995       & 0.576         & -2.740         & -1.492         & 2.471     & -14.621       \\
		\textbf{HEAVY}      &  & -            & -             & -              & -              & -           & -             \\
		\textcolor{black}{\textbf{Real-GARCH}} & & \textcolor{black}{-0.723} & \textcolor{black}{1.729} & \textcolor{black}{-2.397} & \textcolor{black}{-0.231} & \textcolor{black}{2.083} & \textcolor{black}{3.808} \\ 
		\textcolor{black}{\textbf{Real-EGARCH}} & & \textcolor{black}{-2.651} & \textcolor{black}{1.328} & \textcolor{black}{-2.397} & \textcolor{black}{-0.655} & \textcolor{black}{1.855} & \textcolor{black}{1.293} \\ 
		\textcolor{black}{\textbf{ARFIMA}} & & \textcolor{black}{-1.205} &  \textcolor{black}{-1.035} & \textcolor{black}{-1.370} & \textcolor{black}{-0.381} & \textcolor{black}{0.695} & \textcolor{black}{2.550} \\ 
		\textcolor{black}{\textbf{HAR}} & & \textcolor{black}{-0.482} &  \textcolor{black}{-0.654} & \textcolor{black}{-0.342} & \textcolor{black}{-0.047} & \textcolor{black}{0.398} & \textcolor{black}{3.690} \\ 
		\textcolor{black}{\textbf{MLP}} & & \textcolor{black}{-5.586} &  \textcolor{black}{-3.594} & \textcolor{black}{-5.137} & \textcolor{black}{-2.185} & \textcolor{black}{-1.240} & \textcolor{black}{-15.486} \\ 
		\textcolor{black}{\textbf{LSTM}} & & \textcolor{black}{2.716} &  \textcolor{black}{2.022} & \textcolor{black}{0.685} & \textcolor{black}{-0.322} & \textcolor{black}{0.024} & \textcolor{black}{-1.733} \\ 
		\textcolor{black}{\textbf{DeepVol-RV}} & & \textcolor{black}{3.987} & \textcolor{black}{4.062} & \textcolor{black}{2.740} & \textcolor{black}{0.308} & \textcolor{black}{0.536} & \textcolor{black}{5.497}\\ 
		\textbf{DeepVol}    &  & \textbf{14.502}       & \textbf{17.404}        & \textbf{4.452}          & \textbf{0.789}          & 0.863      & 21.097        \\ \thickhline
	\end{tabular}
\end{table}

Table \ref{table:baselines} summarises the out-of-sample forecast performance of the different models we evaluate.
\textcolor{black}{
	Notice that, in addition to the baselines included in Section \ref{subsec:baselines}, we introduce a variation of the proposed model, termed DeepVol-RM. Inspired by the HAR model \citep{corsi2009simple}, DeepVol-RM utilises realised measures computed over the previous twenty-two days as input instead of high-frequency data.
	The primary objective of incorporating this variation of DeepVol is to empirically verify whether there may exist predictive information in the high-frequency data that \textcolor{black}{might be} overlooked when computing a realised measure. We doing so by comparing the prediction accuracy of a  DCC-based model fed with raw high-frequency data (DeepVol) against a DCC-based model that receives realised volatility as input (DeepVol-RM).
} It can be seen that the proposed architecture, DeepVol, improves \textcolor{black}{upon} the baseline results for most metrics, with the exception of ME and MedAE.
Peculiarly, for the later, the martingale process proves to provide the best results. Considering that the MadAE is an outlier-robust error function, this behaviour is not surprising, as the martingale process is the most conservative among the evaluated strategies. For this particular error metric, DeepVol is the second-best in performance terms. \\

As mentioned before, the evaluated baseline methodologies operate in a recurrent manner, utilising  past data, while DeepVol uses just the previous day's intraday information for the day-ahead prediction. Considering these facts, DeepVol's good performance with respect to the MedAE is especially surprising, as noisier behaviour could be expected due to the lack of recurrence.
Furthermore, some of the baseline models evaluated, as the HEAVY model, integrate momentum indicators into their architecture, something that we do not explicitly model in DeepVol.
Consequently, DeepVol's accurate predictions in terms of MAE and RMSE are particularly interesting considering how the model maintains a low MedAE. In conclusion, the proposed architecture shows robustness in the presence of volatility shocks and avoids an escalation on the ME and MedAE as unstable methods would report.\\

\textcolor{black}{
	Concerning DeepVol-RM, which leverages the DeepVol architecture but utilises realised measures from previous days instead of intraday data, it reports the second best results, outperforming the HAR model, from which it took inspiration, and the LSTM and MLP models, which directly utilise intraday data. Its superior performance over HAR is not surprising, as DeepVol-RM benefits from weighting the importance of all realised measures from the previous twenty-two days at inference time, whereas HAR can just aggregates this information, as reflected in Eq. (\ref{eq:har_rm}). This observation is consistent with the findings reported in \citet{reisenhofer2022harnet}, where a similar implementation of a dilated causal convolutions-based network is discussed. Additionally, DeepVol's superior performance compared  to the LSTM model suggests that DCC-based models are more effective than recurrent ones in handling high-frequency data for volatility forecasting.
}\\

Table \ref{table:improvement} extends the results  of Table \ref{table:baselines}, displaying the improvement/degradation for each evaluated method relative to a basic martingale process and the HEAVY model. For the former, we aim to report how much improvement each model provides over the most basic modelling of the problem.  
\textcolor{black}{For the latter, and considering that the HEAVY model is the best performer among the classic models used as baselines, slightly outperforming the HAR model ---as also found in \citet{vis2017volatility}, where they conclude that HEAVY models are generally superior to HAR-type models---, a direct comparison with it is particularly insightful for analyzing DeepVol's performance.}
It can be seen that DeepVol thoroughly overperforms HEAVY concerning the MAE, RMSE, SMAPE, and MedAE errors, while the differences in QLIKE and ME are tighter. As previously mentioned, we consider particularly interesting DeepVol' ability to overperform the rest of the models while proving a robust noise behaviour, avoiding an escalation in the ME and MedAE while performing more accurate predictions.

\subsection{\textcolor{black}{Robustness} Study}
\label{subsec:linearity}


The analysis \textcolor{black}{in Section \ref{subsec:cr}} has \textcolor{black}{demonstrated} that \textcolor{black}{employing} a receptive field of one day and a sampling frequency of 5-minutes \textcolor{black}{yields} the most accurate results for forecasting day-ahead realised volatility. \textcolor{black}{Moreover, we aim to explore potential benefits of incorporating} realised measures into our methodology. \textcolor{black}{The rationale behind this notion} is that integrating \textcolor{black}{information from realised measures of previous days as additional} input would \textcolor{black}{enable} DeepVol to observe a \textcolor{black}{broader} window of past data, \textcolor{black}{thereby complementing} high-frequency data with \textcolor{black}{additional} historical information \textcolor{black}{from} the time series.
\textcolor{black}{Note that this study differs from the variation of DeepVol previously introduced, DeepVol-RM, which forecasts volatility solely based on realised measures from previous days, inspired by the approach proposed in \citet{corsi2009simple}, and does not directly use intraday data as raw input to the DCC-based model.}\\


To \textcolor{black}{incorporate} the realised measures into DeepVol, \textcolor{black}{we make a slight adjustment to its architecture by} adding a linear output as a final layer. \textcolor{black}{This layer combines the} results of the dilated convolutions performed over the high-frequency data\textcolor{black}{, which has a sampling frequency of 5-minutes and a receptive field of 1 day, with the realised measures from the previous $N$ days. The various values of $N$ evaluated can be seen in the column labeled ``Realised Measures Window'' in Table \ref{table:linearity}. Each component of these two is weighted by its corresponding terms.}  \\

The reported results are shown in Table \ref{table:linearity}, where it can be noted that our DNNs-based proposal does not benefit from the inclusion of realised measures as an \textcolor{black}{additional} input feature. Adding the past realised measures results in an analogous behaviour to increasing the receptive field, highlighting again that DeepVol is especially efficient in utilising recent high-frequency data for volatility forecasting, not requiring a more extended lookback window to do so.

\begin{table}[t!]
	\fontsize{10.0}{16}\selectfont
	\centering
	\caption{\textcolor{black}{Robustness} study. DeepVol + RV merges DeepVol's predictions with the realised variance through a linear layer and additional non-linearities.}
	\vspace*{3mm}
	\label{table:linearity}
	\begin{tabular}{cclcccccc}
		\thickhline
		\multicolumn{1}{l}{}                                                          & \multicolumn{1}{l}{\begin{tabular}[c]{@{}c@{}}\textbf{\textcolor{black}{Realised}}\\ \textbf{\textcolor{black}{Measures}}\\ \textbf{\textcolor{black}{Window}}\end{tabular}} &  & \multicolumn{1}{l}{\textbf{MAE}} & \multicolumn{1}{l}{\textbf{RMSE}} & \multicolumn{1}{l}{\textbf{SMAPE}} & \multicolumn{1}{l}{\textbf{QLIKE}} & \multicolumn{1}{l}{\textbf{ME}} & \multicolumn{1}{l}{\textbf{MedAE}} \\ \thickhline
		\multirow{3}{*}{\begin{tabular}[c]{@{}c@{}}\textbf{DeepVol}\\\textbf{+}\\\textbf{RV}\end{tabular}}                                              & \textbf{1}                               &  & 4.130                   & 8.720                    & 0.286                     & 345.150                  & 72.292                 & 2.193                     \\
		& \textbf{2}                               &  & 6.804                   & 10.036                   & 0.335                     & 369.529                   & 69.309                 & 5.728                     \\
		& \textbf{3}                               &  & 6.899                   & 10.008                   & 0.340                     & 371.762                   & 69.903                 & 6.577                     \\ \hline
		\multirow{3}{*}{\begin{tabular}[c]{@{}c@{}}\textbf{DeepVol}\end{tabular}} & \textbf{1}                               &  & \textbf{3.903}                   & 8.457                    & \textbf{0.279}                     & \textbf{340.779}                   & 71.779                 & 2.008                     \\
		& \textbf{2}                               &  & 4.429                   & 9.495                    & 0.308                     & 592.600                   & 78.690                 & \textbf{1.756}                     \\
		& \textbf{3}                               &  & 4.054                   & \textbf{8.379}                    & 0.285                     & 343.359                   & \textbf{70.457}                 & 2.334                     \\ \thickhline
	\end{tabular}
\end{table}

\subsection{Generalisation and Transfer Learning Analysis} 
\label{subsec:outofsample}

Previous experiments used all NASDAQ-100 tickers during training, \textcolor{black}{validation, and} testing, preserving a portion of the dataset's dates for the out-of-sample forecast. In this section, in addition, we split the dataset into two folds in the cross-section. During training \textcolor{black}{and validation}, just half of the tickers are used, while the other half is utilised for testing. For training purposes, we use the first four months of data corresponding to the first half of tickers, that is, from September 30, 2019, through January 31, 2020. \textcolor{black}{Further, we keep the last of these months, that is, January 01 to January 31, 2020, as validation set for hyperparameters selection.} The model is later tested on the remainder tickers, using data from February 01, 2020, through September 30, 2021. This set-up allows us to evaluate the quality of the models' forecasts during the volatility shocks provoked by the COVID-19 crisis, which started in February, 2020.
\textcolor{black}{In} this scenario, \textcolor{black}{we can} evaluate our model's generalisation capabilities, predicting the day-ahead volatility for tickers that were not previously available. \textcolor{black}{As done in Section \ref{subsec:nasdaq}, a 5-minutes sampling frequency and a receptive field of one day are used after validating their values analogously to Section \ref{subsec:cr}.}
The results of this out-of-sample-stocks forecast study are collected in Table \ref{table:out_of_sample}. \\

\begin{table}[t!]
	\fontsize{10.0}{15}\selectfont
	\centering
	\caption{Out-of-sample-stocks forecast. Generalisation Study: experiments results for the NASDAQ-100 dataset. An asterisk ($^*$) indicates models that are included in the MCS at the 5\% significance level.}
	\vspace*{3mm}
	\label{table:out_of_sample}
	\begin{tabular}{cccccccc}
		\thickhline
		\textbf{Method}  & \textbf{} & \textbf{MAE}   & \textbf{RMSE}   & \textbf{SMAPE} & \textbf{QLIKE}    & \textbf{ME} & \textbf{MedAE} \\ \thickhline
		\textbf{martingale}       & \textbf{} & 9.673          & 35.235          & 0.324          & 2142.795          & 341.457           & 2.169 \\
		\textbf{TARCH}   &           & 8.525          & 28.178          & 0.295          & 893.795          & 282.096           & 3.236          \\
		\textbf{IGARCH}  &           & 9.208          & 27.753          & 0.312          & 947.409          & 279.080           & 3.982          \\
		\textbf{FIGARCH} &           & 7.805          & 26.752          & 0.299         & 899.955         & 267.533           & 3.581          \\
		\textbf{APARCH}  &           & 8.179          & 26.749          & 0.297          & 896.063          & \textbf{265.910}$^*$           & 3.557          \\
		\textbf{AGARCH}  &           & 7.928          & 26.486          & 0.295         & 893.682          & 269.577          & 3.191          \\
		\textbf{EGARCH}  &           & 8.180         & 26.767         & 0.297        & 897.432          & 277.022           & 3.530        \\
		\textbf{HEAVY}   &           & 8.315         & 26.322         & 0.294$^*$ & \textbf{874.409}$^*$ & 277.780          & 2.158         \\
		\textcolor{black}{\textbf{Real-GARCH}}      &           & \textcolor{black}{8.402}          & \textcolor{black}{26.523}         & \textcolor{black}{0.297}          & \textcolor{black}{890.894}          & \textcolor{black}{890.894}         & \textcolor{black}{2.301}    \\
		\textcolor{black}{\textbf{Real-EGARCH}}    &           & \textcolor{black}{8.231}          & \textcolor{black}{26.102}         & \textcolor{black}{0.296}          & \textcolor{black}{888.909}          & \textcolor{black}{279.693}         & \textcolor{black}{2.341}    \\
		\textcolor{black}{\textbf{ARFIMA}} & & \textcolor{black}{7.993} & \textcolor{black}{25.001} & \textcolor{black}{0.295} & \textcolor{black}{888.031} & \textcolor{black}{278.882} & \textcolor{black}{2.446} \\
		\textcolor{black}{\textbf{HAR}} & & \textcolor{black}{8.184} & \textcolor{black}{25.042} & \textcolor{black}{0.294}$^*$ & \textcolor{black}{889.471} & \textcolor{black}{279.302} & \textcolor{black}{2.510} \\
		\textcolor{black}{\textbf{MLP}} & & \textcolor{black}{8.902} & \textcolor{black}{27.636} & \textcolor{black}{0.305} & \textcolor{black}{902.342} & \textcolor{black}{289.023} & \textcolor{black}{2.790} \\
		\textcolor{black}{\textbf{LSTM}} & & \textcolor{black}{7.998} & \textcolor{black}{26.941} & \textcolor{black}{0.296} & \textcolor{black}{898.907} & \textcolor{black}{898.907} & \textcolor{black}{2.494} \\
		\textcolor{black}{\textbf{DeepVol-RV}} & & \textcolor{black}{8.514} & \textcolor{black}{26.832} & \textcolor{black}{0.296} & \textcolor{black}{889.880} & \textcolor{black}{282.271} & \textcolor{black}{2.583}\\ 
		\textbf{DeepVol} &           & \textbf{7.288}$^*$ & \textbf{23.396}$^*$ & \textbf{0.292}$^*$          & 894.283         & 275.255     & \textbf{1.927}$^*$        \\ \thickhline
	\end{tabular}
\end{table}

In these conditions, DeepVol still report the best MAE, RMSE, and SMAPE results, while the HEAVY model reports a better QLIKE than the rest of the evaluated methods. Concerning the MeadAE, DeepVol reports the best results, immediately followed by the martingale process, which still outperforms the rest of the baseline models. \textcolor{black}{Furthermore, DeepVol-RM's performance has worsened in comparison to Section \ref{subsec:nasdaq}, presumably due to the fact that in this new environment of volatility shocks, past information contained in the realised measures used as input is less informative}. These results, which are \textcolor{black}{in general} similar to the obtained \textcolor{black}{ones} in the out-of-sample forecast study of Section \ref{subsec:outofsample}, confirm that DeepVol still shows a conservative behaviour in this new forecast scenario, proving its generalisation capabilities to transfer learning from training to test, \textcolor{black}{presumably} learning global features of the data that allow the model to perform well on out-of-distribution data.
As in Section \ref{subsec:nasdaq}, Table \ref{table:improvement_generalisation} reports the improvement/degradation for each evaluated method with respect to a basic martingale process and the HEAVY model on the test set tickers. \\

\begin{table}[t!]
	\fontsize{10.0}{15}\selectfont
	\centering
	\caption{Out-of-sample-stocks forecast: percentage of improvement/degradation over the  martingale process and the HEAVY model. for each of the evaluated models.}

	\vspace*{3mm}
	\label{table:improvement_generalisation}
	\begin{tabular}{clcccccc}
		\thickhline
		\textbf{Method}     &  & \textbf{MAE} & \textbf{RMSE} & \textbf{SMAPE} & \textbf{QLIKE} & \textbf{ME} & \textbf{MedAE} \\ \thickhline
		\textbf{}           &  & \multicolumn{6}{c}{\textbf{Improvement over martingale (\%)}}                                \\ \cline{3-8} 
		\textbf{martingale} &  & -            & -             & -              & -              & -           & -             \\
		\textbf{TARCH}      &  & 11.863       & 20.030        & 9.068         & 58.288         & 17.385       & -49.161     \\
		\textbf{IGARCH}     &  & 4.804       & 21.235         & 3.671         & 55.786         & 18.268      & -83.565      \\
		\textbf{FIGARCH}    &  & 19.315       & 24.075        & 7.711         & 58.001         & 21.650      & -65.089      \\
		\textbf{APARCH}     &  & 15.440       & 24.083       & 8.267         & 58.183         & \textbf{22.125}     & -63.973     \\
		\textbf{AGARCH}     &  & 18.043       & 24.829        & 9.066        & 58.294         & 21.051       & -47.105      \\
		\textbf{EGARCH}     &  &  15.443      & 24.032       & 8.452         & 58.119        & 18.871    & -62.737     \\
		\textbf{HEAVY}      &  & 14.037      & 25.295       & 9.200        & \textbf{59.193}         & 18.649     & 0.534
		\\
		\textcolor{black}{\textbf{Real-GARCH}}      &           & \textcolor{black}{13.138}          & \textcolor{black}{24.725}         & \textcolor{black}{8.390}          & \textcolor{black}{58.424}          & \textcolor{black}{17.994}         & \textcolor{black}{-6.076}    \\
		\textcolor{black}{\textbf{Real-EGARCH}}    &           & \textcolor{black}{14.906}          & \textcolor{black}{25.920}         & \textcolor{black}{8.698}          & \textcolor{black}{58.516}          & \textcolor{black}{18.088}         & \textcolor{black}{-7.920}    \\
		\textcolor{black}{\textbf{ARFIMA}} & & \textcolor{black}{17.366} & \textcolor{black}{29.045} & \textcolor{black}{9.007} & \textcolor{black}{58.557} & \textcolor{black}{18.326} & \textcolor{black}{-12.760} \\
		\textcolor{black}{\textbf{HAR}} & & \textcolor{black}{15.392} & \textcolor{black}{28.928} & \textcolor{black}{9.315} & \textcolor{black}{58.490} & \textcolor{black}{18.203} & \textcolor{black}{-15.711} \\
		\textcolor{black}{\textbf{MLP}} & & \textcolor{black}{7.969} & \textcolor{black}{21.566} & \textcolor{black}{5.922} & \textcolor{black}{57.889} & \textcolor{black}{15.356} & \textcolor{black}{-28.619} \\
		\textcolor{black}{\textbf{LSTM}} & & \textcolor{black}{17.315} & \textcolor{black}{23.539} & \textcolor{black}{8.698} & \textcolor{black}{58.050} & \textcolor{black}{16.976} & \textcolor{black}{-14.973} \\
		\textcolor{black}{\textbf{DeepVol-RV}} & & \textcolor{black}{11.980} & \textcolor{black}{23.850} & \textcolor{black}{8.698} & \textcolor{black}{58.471} & \textcolor{black}{17.333} & \textcolor{black}{-19.076}\\ 
		\textbf{DeepVol}    &  & \textbf{24.660}       & \textbf{33.599}        & \textbf{9.932}         & 58.266        & 19.388      & \textbf{11.165}       \\ \thickhline
		\textbf{}           &  & \multicolumn{6}{c}{\textbf{Improvement over HEAVY (\%)}}                                     \\ \cline{3-8} 
		\textbf{martingale} &  & -16.330      & -33.859       & -10.132       & -145.056     & -22.924    & -0.537        \\
		\textbf{TARCH}      &  &-2.529    & -7.048       & -0.145       & -2.217       & -1.554     & -49.962     \\
		\textbf{IGARCH}     &  & -10.741      & -5.434       &-6.090         & -8.349        & -0.468    & -84.551       \\
		\textbf{FIGARCH}    &  & 6.140     & -1.632       & -1.640        & -2.921         & 3.689     & -65.975         \\
		\textbf{APARCH}     &  & 1.632      & -1.622        & -1.028        & -2.476         & \textbf{4.273}     & -64.854      \\
		\textbf{AGARCH}     &  &4.660       &-0.623        & -0.148       & {-2.204}        & 2.953     & -47.895      \\
		\textbf{EGARCH}     &  & 1.624      & -1.691        & -0.824       & -2.633         & 0.273     & -63.611      \\
		\textbf{HEAVY}      &  & -            & -             & -              & -              & -           & -             \\
		\textcolor{black}{\textbf{Real-GARCH}}      &           & \textcolor{black}{-1.046}          & \textcolor{black}{-0.762}         & \textcolor{black}{-0.892}          & \textcolor{black}{-1.885}          & \textcolor{black}{-0.805}         & \textcolor{black}{-6.646}    \\
		\textcolor{black}{\textbf{Real-EGARCH}}    &           & \textcolor{black}{1.010}          & \textcolor{black}{0.837}         & \textcolor{black}{-0.552}          & \textcolor{black}{-1.658}          & \textcolor{black}{-0.689}         & \textcolor{black}{-8.500}    \\
		\textcolor{black}{\textbf{ARFIMA}} & & \textcolor{black}{3.873} & \textcolor{black}{5.020} & \textcolor{black}{-0.213} & \textcolor{black}{\textbf{-1.558}} & \textcolor{black}{-0.397} & \textcolor{black}{-13.366} \\
		\textcolor{black}{\textbf{HAR}} & & \textcolor{black}{1.575} & \textcolor{black}{4.864} & \textcolor{black}{0.127} & \textcolor{black}{-1.723} & \textcolor{black}{-0.548} & \textcolor{black}{-16.332} \\
		\textcolor{black}{\textbf{MLP}} & & \textcolor{black}{-7.060} & \textcolor{black}{-4.991} & \textcolor{black}{-3.610} & \textcolor{black}{-3.195} & \textcolor{black}{-4.047} & \textcolor{black}{-29.310} \\
		\textcolor{black}{\textbf{LSTM}} & & \textcolor{black}{3.812} & \textcolor{black}{-2.350} & \textcolor{black}{-0.552} & \textcolor{black}{-2.802} & \textcolor{black}{-2.056} & \textcolor{black}{-15.591} \\
		\textcolor{black}{\textbf{DeepVol-RV}} & & \textcolor{black}{-2.393} & \textcolor{black}{-1.936} & \textcolor{black}{-0.552} & \textcolor{black}{-1.769} & \textcolor{black}{-1.617} & \textcolor{black}{-19.716}\\ 
		\textbf{DeepVol}    &  & \textbf{12.357}       & \textbf{11.116}        & \textbf{0.806}          & -2.273          &0.909      & \textbf{10.688}        \\ \thickhline
	\end{tabular}
\end{table}

\begin{figure}[t!]
	\centering
	\begin{minipage}[b]{0.49\textwidth}
		\includegraphics[width=\textwidth]{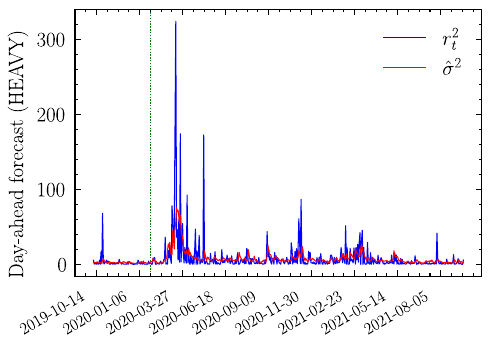}
	\end{minipage}
	\begin{minipage}[b]{0.49\textwidth}
		\includegraphics[width=\textwidth]{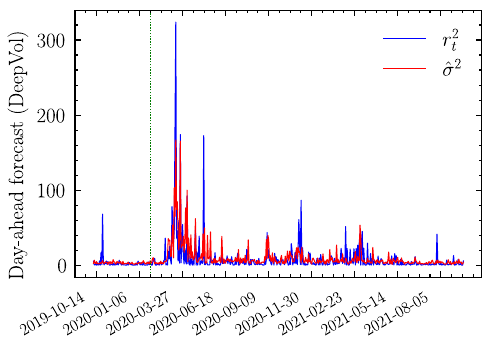}
	\end{minipage}
	\caption{Out-of-sample-stocks: HEAVY's and DeepVol's forecast on PYPL. Green dotted vertical lines mark the forecast start.}
	\label{fig:pypl}
\end{figure}

\begin{figure}[t!]
	\centering
	\begin{minipage}[b]{0.49\textwidth}
		\includegraphics[width=\textwidth]{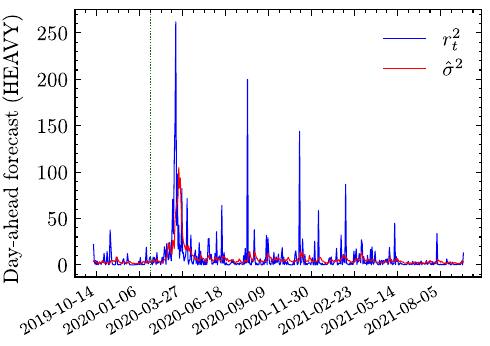}
	\end{minipage}
	\begin{minipage}[b]{0.49\textwidth}
		\includegraphics[width=\textwidth]{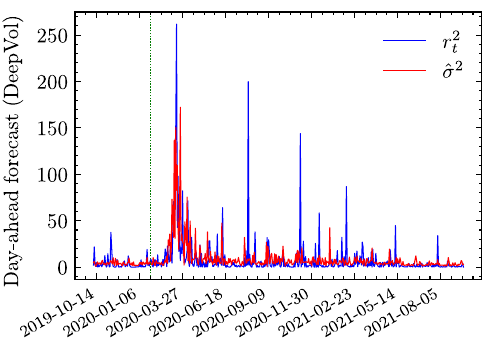}
	\end{minipage}
	\caption{Out-of-sample-stocks: HEAVY's and DeepVol's forecast on QCOM. Green dotted vertical lines mark the forecast start.}
	\label{fig:qcom}
\end{figure}

\begin{figure}[t]
	\centering
	\begin{minipage}[b]{0.49\textwidth}
		\includegraphics[width=\textwidth]{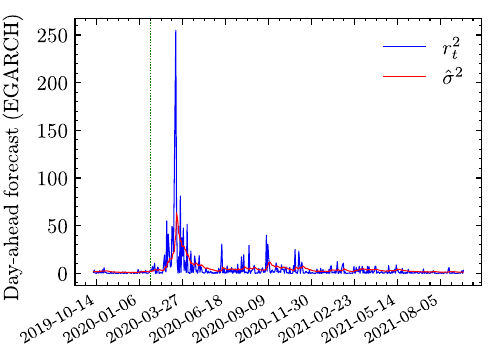}
	\end{minipage}
	\begin{minipage}[b]{0.49\textwidth}
		\includegraphics[width=\textwidth]{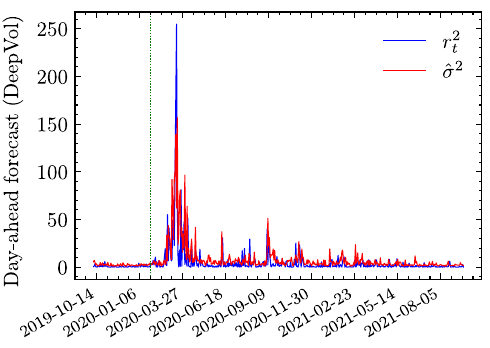}
	\end{minipage}
	\caption{Out-of-sample-stocks: EGARCH's and DeepVol's forecast on MSFT. Green dotted vertical lines mark the forecast start.}
	\label{fig:msft}
\end{figure}

\begin{figure}[t]
	\centering
	\begin{minipage}[b]{0.49\textwidth}
		\includegraphics[width=\textwidth]{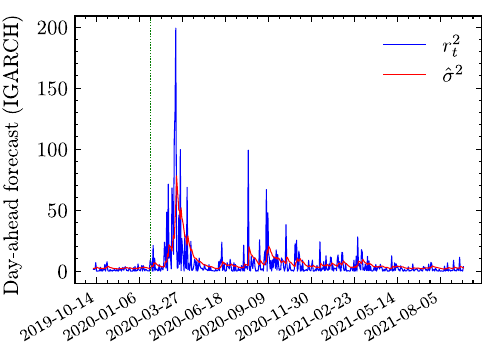}
	\end{minipage}
	\begin{minipage}[b]{0.49\textwidth}
		\includegraphics[width=\textwidth]{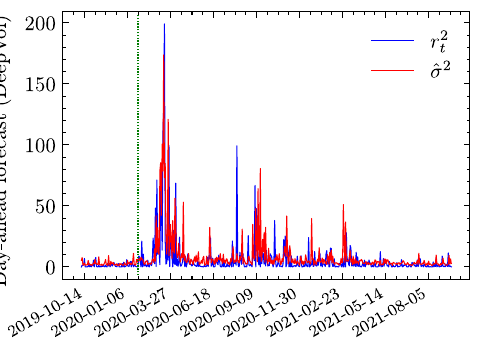}
	\end{minipage}
	\caption{Out-of-sample-stocks: IGARCH's and DeepVol's forecast on AAPL. Green dotted vertical lines mark the forecast start.}
	\label{fig:aapl}
\end{figure}

Finally, Figures \ref{fig:pypl} to \ref{fig:aapl} \textcolor{black}{exemplify how the evaluated models} generalise and transfer learning from the \textcolor{black}{training} tickers \textcolor{black}{to} the test distribution. Model forecasts are \textcolor{black}{juxtaposed} with the daily squared returns, allowing a direct comparison between forecasts from DeepVol and \textcolor{black}{the} baselines. Note that classical methodologies \textcolor{black}{yield} smoother predictions, a phenomenon especially visible in the HEAVY model as it integrates a momentum indicator. 
\textcolor{black}{This conservative behavior, while advantageous for stable conditions, clearly poses a disadvantage}  in terms of slower adaptation to volatility shocks\textcolor{black}{, as seen in the sharp movements during the COVID-19 crisis in 2020 and 2021, evident in the associated figures.}
All the evaluated models reacted to the bigger of these shocks, the 2020 stock market crash, starting in February 2020, in one way or another. 
\textcolor{black}{However}, during the minor shocks that followed that year, baseline predictions are almost negligible with the exception of IGARCH in Fig. \ref{fig:aapl}. HEAVY and EGARCH exhibit an invariable behaviour in this turbulent environment, showing a lack of adaptability to changing conditions.\\

\textcolor{black}{Finally, } we should remark that DeepVol requires just one day of intraday data to perform the out-of-sample volatility forecasting, unlike classical methodologies \textcolor{black}{that} operate recursively \textcolor{black}{and require a} sufficiently long window of past data. This places DeepVol in an advantaged position in situations of low data availability, such as the inclusion of new tickers in the stock market, as it does not require a long horizon of historical to perform its predictions.


\section{Conclusions}
\label{sec:conclusions}

In this paper, we propose a deep learning model based on hierarchies of Dilated Causal Convolutions ---termed DeepVol--- to forecast day-ahead realised volatility from high-frequency data. Our model takes advantage of the automatic feature extraction inherent to Deep Neural Networks to bypass the estimation of the realised measures, tackling the problem of volatility forecasting from a pure data-driven perspective. 
\textcolor{black}{Moreover}, the usage of dilated convolutions enables DeepVol to exponentially increase its input window 
\textcolor{black}{without a significant increase in the number of parameters.} 

\textcolor{black}{While DeepVol's operation does not directly produce any type of realised measure ---the model does not explicitly construct an ex-post estimate of the returns variation--- its aggregation of high-frequency data resembles how realised measures condense intraday data into daily statistics.}
Reported results show how DeepVol's predictions significantly improve the baseline models performance, proving that the proposed data-driven approach avoids the limitations of classical methods, such as model misspecification or the usage of hand-crafted noisy realised measures, by taking advantage of the abundance of high-frequency data.\\

\textcolor{black}{
	Several findings from the experiments are worth highlighting with regard to the usage of Dilated Causal Convolutions for the day-ahead realised volatility forecasting. Firstly, DeepVol generally outperforms traditional autoregressive architectures. 
	Notably, DeepVol's performance remains robust even in high volatility regimes, as evidenced by the reported Median Absolute Error (MedAE) and Mean Error (ME) metrics in Section \ref{sec:experiments}, which indicate quick adaptation to volatility shocks while maintaining some conservatism in their presence.
	These results align with existing literature, indicating that deep-learning-based volatility forecasting architectures \citep{ramos2021multi, zhang2022volatility} and hybrid models \citep{baek2018modaugnet, kim2018forecasting} consistently outperform classical methodologies.
	Further, the comparison between DeepVol-RM, which exclusively utilises the realised variance of previous days as input, and DeepVol itself, sheds more light on why DeepVol outperforms realised-measures-based models in the considered market regimes, suggesting that the model operating on intraday data is capable of leveraging information from the high-frequency returns that would be lost if realised measures were directly used.
}
\textcolor{black}{
	Lastly, our analysis of the receptive field size and sampling frequency in Section \ref{subsec:cr}  reveals that, for the proposed method and in the analysed market conditions, data from the previous trading day contains enough information for predicting the day-ahead realised volatility with high accuracy and therefore, increasing the receptive field of DeepVol does not generally lead to better performance. Further, we find that, in this market regime and for the proxy measure of the volatility employed, a sampling frequency of 5-minutes strikes a balance between capturing intraday information and minimising noise, echoing previous studies analysing this same trade-off in the context of estimation of realised measures from  high-frequency data.}\\

\textcolor{black}{
	Finally, consideration should be given to tailoring all hyperparameters to specific stocks. For instance, adjusting parameters such as the receptive field size might facilitate more conservative predictions if desired in certain contexts, and tailoring the optimal sampling frequency of the intraday data could significantly enhance the model's accuracy. However, in this paper, our aim is to demonstrate the efficacy of a single overarching model capable of learning from a diverse set of stocks and generalizing and transferring learned insights to new stocks. Therefore, practitioners should consider fine-tuning the model's parameters per stock depending on the context.
}\\



\textcolor{black}{In summary, DeepVol offers improved forecasting accuracy compared to baseline methods, proving that the model quickly adapts to volatility shocks while demonstrating noise robustness.} These results are especially relevant
considering that experiments were conducted in high volatility regimes, such as the 2020 stock crisis caused by the  COVID-19 pandemic\textcolor{black}{, a context in which DeepVol shows its ability to extract universal features and transfer learning to out-of-distribution data during the out-of-sample-stocks forecasts.}\\




The empirical results collected in this paper suggest that models based on Dilated Causal Convolutions should be carefully considered in the context of volatility forecasting and as a result can play a key role in the valuation of  financial derivatives, risk management, and portfolio construction.

\section*{Acknowledgements}

The authors would like to thank Álvaro Cartea for his insightful comments, as well as the other members of the Oxford-Man Institute of Quantitative Finance, University of Oxford.

\section*{Funding}

Fernando Moreno-Pino acknowledges support from Spanish government (AEI/MCI) under grants FPU18/00470, RTI2018-099655-B-100, PID2021-123182OB-I00, PID2021-125159NB-I00, and TED2021-131823B-I00, by Comunidad de Madrid under grant IND2022/TIC- 23550, by the European Union (FEDER) and the European Research Council (ERC) through the European Union’s Horizon 2020 research and innovation program under Grant 714161, and by Comunidad de Madrid and FEDER through IntCARE-CM.

\section*{Disclosure of interest}

The authors declare that there are no potential competing interests associated with this article.


\newpage

\appendix

\section{Data Description}\label{app:a}

{\color{black}
	\fontsize{10}{12}\selectfont
	\begin{longtable}{cccccccc}
		\caption{Summary statistics for the NASDAQ stocks' daily returns over the considered time span. }
		\\ \toprule
		\multirow{2}{*}{\textbf{Symbol}} & \multicolumn{7}{c}{\textbf{Daily Returns}}                                                                           \\
		& \textbf{Mean} & \textbf{Std} & \textbf{Median} & \textbf{Minimum} & \textbf{Maximun} & \textbf{Skewness} & \textbf{Kurtosis} \\ \toprule
		\textbf{AAL}   & -0.16                   & 5.79                    & -0.34                   & -29.07                  & 34.56                   & 0.73                    & 8.33                    \\
		\textbf{AAPL}  & 0.27                    & 2.69                    & 0.27                    & -14.11                  & 11.44                   & -0.37                   & 5.17                    \\
		\textbf{ADBE}  & 0.19                    & 2.76                    & 0.38                    & -15.89                  & 16.31                   & -0.15                   & 7.98                    \\
		\textbf{ADI}   & 0.09                    & 2.9                     & 0.13                    & -18.26                  & 13.64                   & -0.59                   & 8.31                    \\
		\textbf{ADP}   & 0.04                    & 2.57                    & 0.16                    & -16.55                  & 11.07                   & -0.75                   & 9.45                    \\
		\textbf{ADSK}  & 0.23                    & 2.94                    & 0.36                    & -12.79                  & 15.05                   & -0.45                   & 5.16                    \\
		\textbf{ALGN}  & 0.34                    & 4.04                    & 0.21                    & -19.35                  & 30.23                   & 1.18                    & 13.17                   \\
		\textbf{ALXN}  & 0.15                    & 2.82                    & -0.04                   & -8.79                   & 25.52                   & 2.37                    & 20.63                   \\
		\textbf{AMAT}  & 0.17                    & 3.53                    & 0.16                    & -22.74                  & 12.06                   & -1.0                    & 7.51                    \\
		\textbf{AMD}   & 0.37                    & 3.56                    & 0.21                    & -15.82                  & 15.24                   & -0.04                   & 3.25                    \\
		\textbf{AMGN}  & 0.06                    & 2.2                     & -0.04                   & -8.71                   & 10.74                   & 0.57                    & 4.82                    \\
		\textbf{AMZN}  & 0.2                     & 2.24                    & 0.21                    & -8.21                   & 7.59                    & -0.03                   & 1.82                    \\
		\textbf{ASML}  & 0.22                    & 2.85                    & 0.19                    & -19.06                  & 11.43                   & -1.1                    & 8.41                    \\
		\textbf{ATVI}  & 0.17                    & 2.28                    & 0.21                    & -11.25                  & 8.64                    & -0.18                   & 3.3                     \\
		\textbf{AVGO}  & 0.16                    & 2.98                    & 0.33                    & -22.2                   & 13.08                   & -1.63                   & 14.23                   \\
		\textbf{BIDU}  & 0.23                    & 2.78                    & 0.06                    & -11.12                  & 12.98                   & 0.59                    & 4.49                    \\
		\textbf{BIIB}  & 0.13                    & 3.6                     & -0.05                   & -10.14                  & 36.53                   & $\sim$                  & $\sim$                  \\
		\textbf{BKNG}  & 0.04                    & 3.0                     & 0.09                    & -11.95                  & 17.4                    & 0.34                    & 6.42                    \\
		\textbf{BMRN}  & 0.09                    & 3.53                    & 0.24                    & -43.58                  & 11.28                   & -6.18                   & 74.46                   \\
		\textbf{CDNS}  & 0.23                    & 2.69                    & 0.44                    & -13.43                  & 12.53                   & -0.28                   & 4.58                    \\
		\textbf{CERN}  & 0.06                    & 1.83                    & 0.14                    & -6.28                   & 6.89                    & 0.02                    & 2.08                    \\
		\textbf{CHKP}  & 0.06                    & 1.99                    & 0.1                     & -10.07                  & 13.2                    & 0.44                    & 9.72                    \\
		\textbf{CHTR}  & 0.14                    & 2.07                    & 0.15                    & -15.53                  & 9.43                    & -0.91                   & 12.57                   \\
		\textbf{CMCSA} & 0.06                    & 2.22                    & 0.11                    & -8.89                   & 12.17                   & 0.17                    & 5.19                    \\
		\textbf{COST}  & 0.09                    & 1.68                    & 0.08                    & -6.66                   & 9.45                    & 0.78                    & 7.63                    \\
		\textbf{CSCO}  & -0.01                   & 2.45                    & 0.08                    & -11.84                  & 12.52                   & -0.39                   & 7.45                    \\
		\textbf{CSX}   & 0.1                     & 2.66                    & 0.19                    & -16.87                  & 15.23                   & -0.54                   & 10.73                   \\
		\textbf{CTAS}  & 0.1                     & 3.0                     & 0.24                    & -18.51                  & 13.72                   & -0.96                   & 9.43                    \\
		\textbf{CTSH}  & 0.1                     & 2.76                    & 0.04                    & -18.9                   & 14.01                   & -0.39                   & 10.71                   \\
		\textbf{CTXS}  & 0.1                     & 2.36                    & 0.25                    & -13.87                  & 14.06                   & -0.04                   & 8.88                    \\
		\textbf{DLTR}  & -0.02                   & 2.91                    & 0.09                    & -17.33                  & 13.16                   & -0.97                   & 12.22                   \\
		\textbf{EA}    & 0.13                    & 2.12                    & 0.21                    & -9.32                   & 10.06                   & -0.15                   & 4.65                    \\
		\textbf{EBAY}  & 0.09                    & 2.26                    & 0.14                    & -9.64                   & 8.36                    & -0.39                   & 2.13                    \\
		\textbf{EXPE}  & 0.0                     & 4.44                    & 0.13                    & -31.79                  & 21.88                   & -1.32                   & 13.8                    \\
		\textbf{FAST}  & 0.15                    & 2.43                    & 0.09                    & -11.75                  & 15.83                   & 0.82                    & 9.48                    \\
		\textbf{FB}    & 0.13                    & 2.68                    & 0.22                    & -15.61                  & 9.95                    & -0.49                   & 5.18                    \\
		\textbf{FISV}  & 0.02                    & 2.72                    & 0.13                    & -18.27                  & 11.81                   & -0.75                   & 8.71                    \\
		\textbf{GILD}  & -0.01                   & 2.18                    & -0.11                   & -9.03                   & 9.27                    & 0.39                    & 3.33                    \\
		\textbf{HAS}   & -0.06                   & 3.3                     & 0.14                    & -20.77                  & 19.61                   & -1.03                   & 12.91                   \\
		\textbf{HSIC}  & 0.02                    & 2.57                    & 0.13                    & -10.14                  & 9.77                    & -0.21                   & 2.77                    \\
		\textbf{IDXX}  & 0.19                    & 2.44                    & 0.4                     & -12.74                  & 11.46                   & -0.4                    & 4.51                    \\
		\textbf{ILMN}  & 0.06                    & 2.74                    & 0.3                     & -11.69                  & 10.19                   & -0.84                   & 3.39                    \\
		\textbf{INCY}  & 0.05                    & 2.49                    & -0.13                   & -9.86                   & 8.37                    & 0.14                    & 1.53                    \\
		\textbf{INTC}  & 0.0                     & 3.13                    & 0.12                    & -20.04                  & 17.87                   & -0.92                   & 12.9                    \\
		\textbf{INTU}  & 0.12                    & 2.73                    & 0.22                    & -15.69                  & 18.3                    & 0.23                    & 10.0                    \\
		\textbf{ISRG}  & 0.14                    & 2.65                    & 0.23                    & -15.32                  & 15.85                   & -0.25                   & 8.53                    \\
		\textbf{JBHT}  & 0.08                    & 2.39                    & 0.13                    & -15.44                  & 12.33                   & -0.63                   & 9.26                    \\
		\textbf{JD}    & 0.35                    & 2.9                     & 0.41                    & -13.5                   & 11.33                   & -0.17                   & 2.55                    \\
		\textbf{KHC}   & 0.1                     & 2.58                    & 0.19                    & -14.51                  & 17.81                   & 0.72                    & 12.62                   \\
		\textbf{KLAC}  & 0.16                    & 3.49                    & 0.13                    & -14.49                  & 15.89                   & 0.15                    & 5.13                    \\
		\textbf{LRCX}  & 0.23                    & 3.75                    & 0.24                    & -20.36                  & 18.39                   & -0.42                   & 6.77                    \\
		\textbf{LULU}  & 0.19                    & 3.03                    & 0.4                     & -24.4                   & 11.46                   & -1.88                   & 15.17                   \\
		\textbf{MAR}   & 0.03                    & 3.78                    & 0.15                    & -16.48                  & 17.81                   & 0.22                    & 4.51                    \\
		\textbf{MCHP}  & 0.13                    & 3.59                    & 0.09                    & -22.67                  & 14.51                   & -0.67                   & 8.33                    \\
		\textbf{MDLZ}  & 0.02                    & 1.96                    & 0.04                    & -12.14                  & 10.78                   & -0.02                   & 11.03                   \\
		\textbf{MELI}  & 0.35                    & 3.44                    & 0.53                    & -13.68                  & 17.9                    & -0.3                    & 4.01                    \\
		\textbf{MNST}  & 0.16                    & 2.06                    & 0.24                    & -8.28                   & 8.39                    & 0.09                    & 4.4                     \\
		\textbf{MSFT}  & 0.16                    & 2.53                    & 0.22                    & -15.96                  & 13.52                   & -0.44                   & 8.76                    \\
		\textbf{MU}    & 0.17                    & 3.44                    & 0.16                    & -22.13                  & 12.3                    & -0.8                    & 6.56                    \\
		\textbf{MYL}   & -0.06                   & 2.82                    & 0.0                     & -10.34                  & 12.87                   & -0.12                   & 2.19                    \\
		\textbf{NFLX}  & 0.22                    & 2.78                    & 0.13                    & -12.48                  & 10.95                   & -0.3                    & 2.88                    \\
		\textbf{NTAP}  & 0.09                    & 3.04                    & 0.2                     & -12.02                  & 12.88                   & -0.18                   & 3.09                    \\
		\textbf{NTES}  & 0.2                     & 2.66                    & 0.12                    & -12.11                  & 8.68                    & -0.3                    & 2.09                    \\
		\textbf{NVDA}  & 0.34                    & 3.39                    & 0.33                    & -20.3                   & 16.22                   & -0.59                   & 6.18                    \\
		\textbf{NXPI}  & 0.12                    & 3.48                    & 0.34                    & -20.47                  & 15.44                   & -0.61                   & 7.62                    \\
		\textbf{ORLY}  & 0.05                    & 2.49                    & 0.01                    & -16.76                  & 12.15                   & -0.95                   & 12.74                   \\
		\textbf{PAYX}  & 0.04                    & 2.76                    & 0.15                    & -21.65                  & 16.48                   & -1.12                   & 17.44                   \\
		\textbf{PCAR}  & 0.09                    & 2.09                    & 0.17                    & -7.95                   & 13.34                   & 0.66                    & 7.34                    \\
		\textbf{PEP}   & 0.03                    & 2.11                    & 0.09                    & -12.13                  & 12.18                   & -0.64                   & 14.88                   \\
		\textbf{PYPL}  & 0.26                    & 3.07                    & 0.21                    & -17.99                  & 13.19                   & -0.1                    & 6.52                    \\
		\textbf{QCOM}  & 0.23                    & 3.07                    & 0.02                    & -16.18                  & 14.14                   & 0.0                     & 5.29                    \\
		\textbf{REGN}  & 0.17                    & 2.54                    & 0.11                    & -10.91                  & 11.27                   & 0.23                    & 2.38                    \\
		\textbf{ROST}  & 0.04                    & 3.3                     & 0.17                    & -22.15                  & 15.67                   & -0.43                   & 10.52                   \\
		\textbf{SBUX}  & 0.08                    & 2.63                    & 0.04                    & -18.17                  & 13.86                   & -0.5                    & 11.51                   \\
		\textbf{SIRI}  & 0.0                     & 2.54                    & 0.15                    & -16.35                  & 9.69                    & -1.21                   & 8.74                    \\
		\textbf{SNPS}  & 0.2                     & 2.5                     & 0.45                    & -14.15                  & 10.91                   & -0.52                   & 5.59                    \\
		\textbf{SWKS}  & 0.21                    & 3.28                    & 0.3                     & -20.07                  & 15.04                   & -0.38                   & 6.72                    \\
		\textbf{TMUS}  & 0.17                    & 2.32                    & 0.28                    & -11.85                  & 11.06                   & -0.13                   & 7.63                    \\
		\textbf{TSLA}  & 0.87                    & 5.22                    & 0.82                    & -23.62                  & 18.19                   & -0.49                   & 3.72                    \\
		\textbf{TTWO}  & 0.17                    & 2.47                    & 0.25                    & -12.51                  & 9.32                    & -0.78                   & 4.46                    \\
		\textbf{TXN}   & 0.09                    & 2.56                    & 0.17                    & -12.78                  & 12.7                    & -0.09                   & 5.05                    \\
		\textbf{UAL}   & -0.21                   & 5.94                    & -0.27                   & -36.31                  & 22.85                   & -0.61                   & 7.27                    \\
		\textbf{ULTA}  & 0.04                    & 3.81                    & 0.09                    & -24.86                  & 21.75                   & -0.1                    & 12.39                   \\
		\textbf{VRSK}  & 0.09                    & 2.22                    & 0.11                    & -11.24                  & 12.67                   & -0.11                   & 8.79                    \\
		\textbf{VRSN}  & 0.05                    & 2.3                     & 0.19                    & -13.05                  & 12.0                    & 0.09                    & 8.0                     \\
		\textbf{VRTX}  & 0.1                     & 2.75                    & 0.1                     & -23.29                  & 10.15                   & -1.9                    & 17.24                   \\
		\textbf{WBA}   & -0.07                   & 2.82                    & -0.14                   & -11.63                  & 11.89                   & -0.08                   & 3.16                    \\
		\textbf{WDAY}  & 0.1                     & 3.12                    & 0.36                    & -13.66                  & 11.86                   & -0.26                   & 3.26                    \\
		\textbf{WDC}   & -0.01                   & 4.21                    & 0.0                     & -22.91                  & 16.34                   & -1.11                   & 5.89                    \\
		\textbf{WLTW}  & 0.04                    & 2.32                    & 0.05                    & -9.66                   & 14.01                   & 0.15                    & 7.75                    \\
		\textbf{WYNN}  & 0.02                    & 4.99                    & -0.21                   & -28.0                   & 24.56                   & -0.38                   & 6.47                    \\
		\textbf{XEL}   & 0.02                    & 2.25                    & 0.13                    & -13.57                  & 10.37                   & -0.83                   & 10.46                   \\
		\textbf{XLNX}  & 0.14                    & 2.86                    & -0.11                   & -11.32                  & 13.19                   & 0.4                     & 3.15                         \\ \bottomrule
	\end{longtable}
	\label{table:daily_returns}
}

\afterpage{
{\color{black}
	\fontsize{10}{12}\selectfont
	\begin{longtable}{cccccccc}
		\caption{Summary statistics for the NASDAQ stocks' daily realised variance over the considered time span.}
		\\ \toprule
		\multirow{2}{*}{\textbf{Symbol}} & \multicolumn{7}{c}{\textbf{Daily Realised  Variance}}                                                                           \\
		& \textbf{Mean} & \textbf{Std} & \textbf{Median} & \textbf{Minimum} & \textbf{Maximun} & \textbf{Skewness} & \textbf{Kurtosis} \\ \toprule
		\textbf{AAL}                     & 22.6          & 42.86        & 10.31           & 1.13         & 411.94       & 4.97          & 31.55             \\
		\textbf{AAPL}                    & 4.04          & 6.83         & 1.83            & 0.26         & 49.23        & 4.06          & 18.58             \\
		\textbf{ADBE}                    & 4.4           & 6.47         & 2.48            & 0.26         & 47.92        & 4.08          & 19.18             \\
		\textbf{ADI}                     & 5.13          & 9.85         & 2.44            & 0.33         & 87.12        & 5.17          & 30.21             \\
		\textbf{ADP}                     & 4.8           & 10.7         & 1.89            & 0.31         & 102.61       & 5.35          & 33.47             \\
		\textbf{ADSK}                    & 6.5           & 9.43         & 3.84            & 0.37         & 85.45        & 4.52          & 25.56             \\
		\textbf{ALGN}                    & 11.24         & 16.42        & 6.36            & 1.72         & 132.13       & 4.41          & 23.6              \\
		\textbf{ALXN}                    & 6.18          & 7.92         & 3.81            & 0.32         & 63.54        & 4.16          & 20.04             \\
		\textbf{AMAT}                    & 5.5           & 8.94         & 2.88            & 0.64         & 66.71        & 4.61          & 23.61             \\
		\textbf{AMD}                     & 7.61          & 8.72         & 4.83            & 1.08         & 64.31        & 3.55          & 14.47             \\
		\textbf{AMGN}                    & 4.39          & 8.34         & 2.13            & 0.38         & 62.02        & 4.79          & 24.59             \\
		\textbf{AMZN}                    & 3.42          & 4.72         & 2.0             & 0.15         & 41.64        & 4.01          & 21.0              \\
		\textbf{ASML}                    & 3.53          & 5.48         & 1.79            & 0.47         & 36.73        & 4.06          & 18.13             \\
		\textbf{ATVI}                    & 4.83          & 7.66         & 2.53            & 0.57         & 67.18        & 4.76          & 26.15             \\
		\textbf{AVGO}                    & 5.22          & 10.89        & 2.22            & 0.48         & 96.23        & 5.02          & 28.45             \\
		\textbf{BIDU}                    & 5.19          & 5.17         & 3.47            & 0.67         & 36.34        & 3.01          & 10.53             \\
		\textbf{BIIB}                    & 7.01          & 19.78        & 2.99            & 0.03         & 302.31       & 11.42         & 159.51            \\
		\textbf{BKNG}                    & 9.13          & 11.49        & 5.76            & 0.68         & 88.17        & 3.66          & 16.04             \\
		\textbf{BMRN}                    & 7.64          & 10.22        & 4.68            & 1.33         & 74.02        & 4.24          & 20.59             \\
		\textbf{CDNS}                    & 6.61          & 24.02        & 2.9             & 0.8          & 403.4        & 14.65         & 236.91            \\
		\textbf{CERN}                    & 4.54          & 8.79         & 1.95            & 0.39         & 69.47        & 4.56          & 23.22             \\
		\textbf{CHKP}                    & 5.45          & 13.48        & 2.34            & 0.59         & 176.49       & 8.44          & 90.08             \\
		\textbf{CHTR}                    & 5.2           & 9.85         & 2.34            & 0.43         & 85.43        & 5.04          & 30.63             \\
		\textbf{CMCSA}                   & 4.35          & 10.02        & 1.76            & 0.44         & 81.18        & 5.2           & 28.69             \\
		\textbf{COST}                    & 2.55          & 5.55         & 1.03            & 0.18         & 44.94        & 4.98          & 27.4              \\
		\textbf{CSCO}                    & 3.63          & 8.27         & 1.45            & 0.35         & 71.13        & 5.23          & 29.96             \\
		\textbf{CSX}                     & 4.68          & 9.75         & 2.03            & 0.36         & 85.58        & 4.88          & 27.26             \\
		\textbf{CTAS}                    & 8.09          & 16.02        & 4.01            & 0.76         & 195.18       & 6.77          & 63.57             \\
		\textbf{CTSH}                    & 10.71         & 106.99       & 2.15            & 0.34         & 1892.0       & 17.45         & 304.39            \\
		\textbf{CTXS}                    & 5.23          & 9.42         & 2.8             & 0.25         & 74.6         & 4.7           & 25.05             \\
		\textbf{DLTR}                    & 5.79          & 12.69        & 2.75            & 0.66         & 119.42       & 6.05          & 42.69             \\
		\textbf{EA}                      & 4.7           & 7.22         & 2.66            & 0.4          & 61.63        & 4.5           & 23.88             \\
		\textbf{EBAY}                    & 4.17          & 5.87         & 2.56            & 0.22         & 48.64        & 4.02          & 19.12             \\
		\textbf{EXPE}                    & 11.83         & 22.11        & 5.94            & 0.63         & 194.6        & 5.33          & 34.43             \\
		\textbf{FAST}                    & 4.98          & 9.67         & 2.1             & 0.57         & 76.47        & 4.46          & 22.05             \\
		\textbf{FB}                      & 4.33          & 6.2          & 2.69            & 0.41         & 52.23        & 4.59          & 24.9              \\
		\textbf{FISV}                    & 5.47          & 13.59        & 2.41            & 0.31         & 191.08       & 9.38          & 113.63            \\
		\textbf{GILD}                    & 4.53          & 10.12        & 1.71            & 0.48         & 86.76        & 5.73          & 37.25             \\
		\textbf{HAS}                     & 7.9           & 12.87        & 4.26            & 0.69         & 92.6         & 4.36          & 20.86             \\
		\textbf{HSIC}                    & 7.83          & 12.86        & 4.32            & 0.5          & 112.02       & 4.53          & 24.85             \\
		\textbf{IDXX}                    & 9.79          & 16.12        & 4.73            & 0.91         & 128.64       & 4.4           & 23.07             \\
		\textbf{ILMN}                    & 6.65          & 10.38        & 3.62            & 0.8          & 77.88        & 4.21          & 19.46             \\
		\textbf{INCY}                    & 9.71          & 56.18        & 4.2             & 1.22         & 986.19       & 16.88         & 290.59            \\
		\textbf{INTC}                    & 4.7           & 11.21        & 1.84            & 0.35         & 99.29        & 5.63          & 35.5              \\
		\textbf{INTU}                    & 5.57          & 9.48         & 2.74            & 0.56         & 81.0         & 4.62          & 24.89             \\
		\textbf{ISRG}                    & 6.33          & 10.29        & 3.4             & 0.8          & 98.68        & 4.98          & 30.94             \\
		\textbf{JBHT}                    & 7.5           & 14.74        & 3.6             & 0.71         & 135.71       & 5.58          & 38.12             \\
		\textbf{JD}                      & 6.3           & 7.6          & 4.08            & 0.67         & 85.1         & 5.18          & 40.26             \\
		\textbf{KHC}                     & 4.4           & 8.79         & 2.05            & 0.47         & 72.58        & 5.37          & 31.64             \\
		\textbf{KLAC}                    & 6.61          & 9.28         & 3.85            & 0.87         & 75.85        & 4.4           & 23.24             \\
		\textbf{LRCX}                    & 6.99          & 9.56         & 3.98            & 0.72         & 70.87        & 3.91          & 17.89             \\
		\textbf{LULU}                    & 6.68          & 11.81        & 3.94            & 0.53         & 139.13       & 6.53          & 57.39             \\
		\textbf{MAR}                     & 13.45         & 51.38        & 4.68            & 0.3          & 715.86       & 10.62         & 128.32            \\
		\textbf{MCHP}                    & 6.6           & 9.93         & 3.65            & 0.45         & 90.88        & 4.54          & 25.74             \\
		\textbf{MDLZ}                    & 3.36          & 8.73         & 1.17            & 0.32         & 81.08        & 5.64          & 35.21             \\
		\textbf{MELI}                    & 13.81         & 14.73        & 9.05            & 1.16         & 106.06       & 3.52          & 15.27             \\
		\textbf{MNST}                    & 4.48          & 9.43         & 1.93            & 0.52         & 99.82        & 5.95          & 43.66             \\
		\textbf{MSFT}                    & 3.76          & 8.12         & 1.43            & 0.09         & 71.93        & 5.14          & 30.36             \\
		\textbf{MU}                      & 6.53          & 11.06        & 3.52            & 0.84         & 97.46        & 5.22          & 31.78             \\
		\textbf{MYL}                     & 10.03         & 15.66        & 5.42            & 1.21         & 114.53       & 4.22          & 19.84             \\
		\textbf{NFLX}                    & 5.91          & 7.47         & 3.41            & 0.8          & 62.96        & 4.29          & 22.29             \\
		\textbf{NTAP}                    & 7.24          & 17.28        & 3.46            & 0.76         & 250.72       & 9.97          & 127.58            \\
		\textbf{NTES}                    & 6.74          & 7.0          & 4.52            & 1.21         & 57.85        & 3.72          & 17.52             \\
		\textbf{NVDA}                    & 6.69          & 10.66        & 3.55            & 0.71         & 105.72       & 4.88          & 31.05             \\
		\textbf{NXPI}                    & 7.54          & 13.29        & 3.83            & 0.75         & 138.6        & 5.85          & 42.58             \\
		\textbf{ORLY}                    & 7.36          & 19.65        & 2.91            & 0.62         & 293.07       & 10.54         & 142.63            \\
		\textbf{PAYX}                    & 5.64          & 15.43        & 1.72            & 0.29         & 148.48       & 5.91          & 40.68             \\
		\textbf{PCAR}                    & 4.69          & 7.75         & 2.19            & 0.59         & 62.82        & 4.16          & 20.28             \\
		\textbf{PEP}                     & 3.42          & 10.31        & 1.12            & 0.16         & 120.15       & 7.23          & 63.74             \\
		\textbf{PYPL}                    & 5.6           & 8.49         & 3.19            & 0.27         & 58.1         & 4.24          & 20.26             \\
		\textbf{QCOM}                    & 5.01          & 7.66         & 2.82            & 0.35         & 60.25        & 4.52          & 22.97             \\
		\textbf{REGN}                    & 7.26          & 9.1          & 4.53            & 1.08         & 72.22        & 4.32          & 22.3              \\
		\textbf{ROST}                    & 6.99          & 13.47        & 3.48            & 0.55         & 109.05       & 5.13          & 30.07             \\
		\textbf{SBUX}                    & 4.38          & 10.26        & 1.51            & 0.17         & 83.66        & 5.06          & 28.21             \\
		\textbf{SIRI}                    & 6.31          & 9.3          & 3.82            & 1.17         & 93.44        & 5.13          & 33.54             \\
		\textbf{SNPS}                    & 6.13          & 8.58         & 3.71            & 0.8          & 59.56        & 4.14          & 19.0              \\
		\textbf{SWKS}                    & 6.35          & 10.77        & 3.44            & 0.92         & 110.83       & 5.41          & 36.91             \\
		\textbf{TMUS}                    & 4.23          & 7.63         & 1.98            & 0.19         & 66.62        & 4.86          & 28.33             \\
		\textbf{TSLA}                    & 16.15         & 22.21        & 8.25            & 0.98         & 152.42       & 3.18          & 11.75             \\
		\textbf{TTWO}                    & 5.99          & 8.52         & 3.63            & 0.84         & 77.57        & 4.82          & 28.95             \\
		\textbf{TXN}                     & 3.96          & 7.48         & 1.86            & 0.24         & 72.12        & 5.29          & 33.65             \\
		\textbf{UAL}                     & 22.93         & 57.24        & 8.2             & 0.57         & 575.89       & 6.52          & 49.71             \\
		\textbf{ULTA}                    & 11.11         & 21.24        & 5.38            & 1.21         & 156.61       & 4.35          & 19.75             \\
		\textbf{VRSK}                    & 10.73         & 73.47        & 2.28            & 0.46         & 1269.28      & 16.21         & 273.97            \\
		\textbf{VRSN}                    & 6.4           & 12.2         & 2.8             & 0.72         & 118.17       & 4.96          & 30.98             \\
		\textbf{VRTX}                    & 9.19          & 47.49        & 3.43            & 0.77         & 823.47       & 16.25         & 275.06            \\
		\textbf{WBA}                     & 5.73          & 12.23        & 3.01            & 0.65         & 155.82       & 7.74          & 78.06             \\
		\textbf{WDAY}                    & 7.21          & 10.28        & 4.71            & 0.94         & 118.09       & 6.03          & 49.52             \\
		\textbf{WDC}                     & 9.03          & 15.67        & 4.52            & 1.46         & 132.72       & 4.78          & 25.64             \\
		\textbf{WLTW}                    & 7.34          & 11.88        & 3.62            & 0.64         & 96.86        & 4.1           & 20.33             \\
		\textbf{WYNN}                    & 16.28         & 35.4         & 7.5             & 1.03         & 360.04       & 6.04          & 43.25             \\
		\textbf{XEL}                     & 4.63          & 12.95        & 1.62            & 0.35         & 170.39       & 8.36          & 90.58             \\
		\textbf{XLNX}                    & 4.94          & 6.77         & 2.97            & 0.74         & 57.79        & 4.69          & 25.99            \\ \bottomrule
	\end{longtable}
	\label{table:daily_realised_variance}
	}}
	
	\clearpage
	\newpage
	\bibliographystyle{tfcad}
	\bibliography{DeepVol}
	
\end{document}